 \documentclass[epjST]{svjour}

\usepackage{graphicx}
\usepackage{lineno}
\usepackage{amssymb}
\usepackage{color}

\usepackage[nowatermark]{fixmetodonotes}

\usepackage{isotope}
\usepackage{siunitx}
\usepackage[
natbib=true,
style=numeric-comp,
sorting=none]{biblatex}
\usepackage{url}

\usepackage{hyperref}

\usepackage{amsmath}
\usepackage{xcolor}

\begin{document}
\title{
\begin{minipage}[c]{0.75\textwidth}
Recent highlights from GENIE v3
\end{minipage}
\begin{minipage}[c]{0.12\textwidth}
\includegraphics[width=0.95\textwidth]{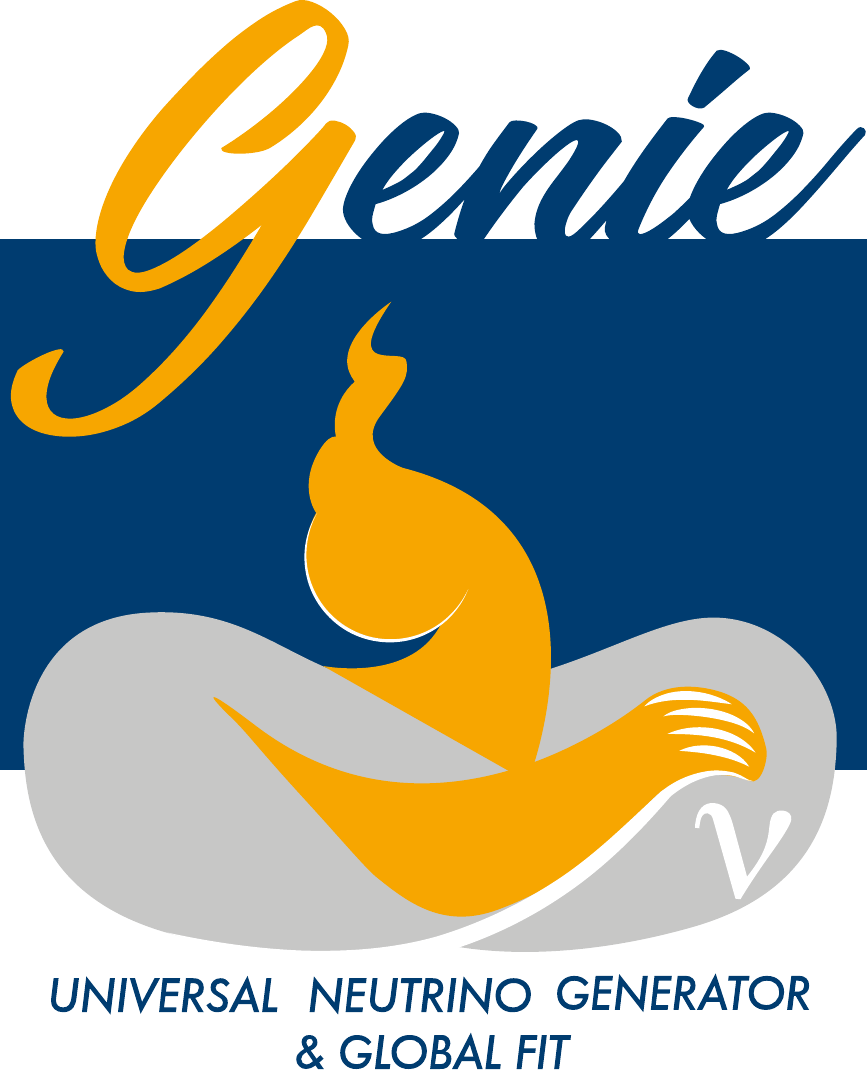}%
\end{minipage}
}
\subtitle{%
}

\author{
    \begin{center}
    {\bf The GENIE Collaboration}\\
    \end{center}
    Luis Alvarez-Ruso\inst{5}
    \and
    Costas Andreopoulos\inst{8,11}
      \fnmsep\thanks{\email{publications@genie-mc.org}}
    \and
    Adi Ashkenazi\inst{9}
      \fnmsep\thanks{Now at Tel Aviv University}
    \and
    Christopher Barry\inst{8}
    \and
    Steve Dennis\inst{8}
      \fnmsep\thanks{Now at University of Cambridge}
    \and
    Steve Dytman\inst{10}
    \and
    Hugh Gallagher\inst{12}
    \and
    Alfonso Andres Garcia Soto \inst{4,5}
    \and
    Steven Gardiner\inst{3}
    \and
    Walter Giele\inst{3}
    \and
    Robert Hatcher\inst{3}
    \and
    Or Hen\inst{9}
    \and
    Libo Jiang\inst{10}
      \fnmsep\thanks{Now at Virginia Polytechnic Institute and State University}
   \and
    Igor D. Kakorin\inst{7}
    \and
    Konstantin S. Kuzmin\inst{6,7}
    \and
    Anselmo Meregaglia\inst{2}
    \and
    Vadim A. Naumov\inst{7}
    \and
    Afroditi Papadopoulou\inst{9}
    \and
    Marco Roda \inst{8}
    \and
    Vladyslav Syrotenko\inst{12}
    \and
    J\'ulia Tena-Vidal\inst{8}
    \and
    Jeremy Wolcott\inst{12}
    \and
    Natalie Wright\inst{9}\\
    \begin{center}
    {\bf Additional authors}\\
    \end{center}
    Monireh Kabirnezhad\inst{13}
    \and
    Narisoa Vololoniaina\inst{1}
}
\institute{
    University of Antananarivo, 
    Dept. of Physics, 
    101 Antananarivo, Madagascar
    \and
    CENBG, Universit\'e de Bordeaux,
    CNRS/IN2P3, 33175 Gradignan, France
    \and
    Fermi National Accelerator Laboratory,
    Batavia, Illinois 60510, USA
    \and
    Harvard University, Dept. of Physics,
    Cambridge, MA 02138, USA
    \and
    Instituto de F\'{i}sica Corpuscular (IFIC),
    Consejo Superior de Investigaciones Cient\'{i}ficas (CSIC) y
    Universitat de Val\`{e}ncia (UV), 
    46980 Paterna, Val\`{e}ncia, Spain
    \and
    Alikhanov Institute for Theoretical and Experimental Physics (ITEP) \\
    of NRC ``Kurchatov Institute'',
    Moscow, 117218, Russia
    \and
    Joint Institute for Nuclear Research (JINR),
    Dubna, Moscow region, 141980, Russia
    \and
    University of Liverpool,
    Dept. of Physics,
    Liverpool L69 7ZE, UK \and
    Massachusetts Institute of Technology,
    Dept. of Physics,
    Cambridge, MA 02139, USA
    \and
    University of Pittsburgh,
    Dept. of Physics and Astronomy,
    Pittsburgh PA 15260, USA
    \and
    U.K. Research and Innovation,
    Science and Technology Facilities Council,
    Rutherford Appleton Laboratory,
    Particle Physics Dept.,
    Oxfordshire OX11 0QX, UK
    \and
    Tufts University,
    Dept. of Physics and Astronomy,
    Medford MA 02155, USA
    \and
    York University, 
    Dept. of Physics and Astronomy, 
    Toronto, ONT M3J 1P3, Canada
}


%
\abstract{
The release of GENIE v3.0.0 was a major milestone in the long history of the GENIE project,
delivering several alternative comprehensive neutrino interaction models, 
improved charged-lepton scattering simulations,
a range of beyond the Standard Model simulation capabilities,
improved experimental interfaces, 
expanded core framework capabilities,
and advanced new frameworks for the global analysis of neutrino scattering data 
and tuning of neutrino interaction models.
Steady progress continued following the release of GENIE v3.0.0.
New tools and a large number of new physics models, comprehensive model configurations, 
and tunes have been made publicly available and planned for release in v3.2.0.
This article highlights some of the most recent technical and physics developments
in the GENIE v3 series.
} 
\maketitle





%
\section{Introduction}
\label{sec:intro}

The release of GENIE v3.0.0 was a major milestone in the long history of the GENIE project \cite{Andreopoulos:2009rq}. The associated technical and physics modelling developments underlined the dual role of GENIE in: a) maintaining the single universal platform for delivering well-validated and state-of-the-art physics simulations directly into the established Monte Carlo (MC) simulation chains of nearly all neutrino experiments, and b) taking the leading role in the development, validation, characterisation and tuning of {\em comprehensive} neutrino simulations, that incorporate descriptions for all relevant processes across the full kinematic space accessible by different types of neutrino experiments.
Addressing the community demand for {\em alternative} models, GENIE v3.0.0 amalgamated large collections of modelling elements, many of which were developed with strong community support, into a number of distinct and relatively consistent comprehensive model configurations that were validated, characterised, tuned and deployed as a whole \cite{free_nucleon_tune}. 
They were the seeds around which many modelling developments have coalesced, leading to more well-motivated variants and tunes. 
The maturation of the collaborative GENIE development paradigm, along with the substantial effort invested in curating extensive data archives of neutrino, electron and hadron scattering data, developing advanced frameworks for data/MC comparisons, tuning, and continuous integration underpinned a marked improvement both in the volume of deployed simulations and in release frequency. 

Steady progress continued following the release of GENIE v3.0.0, with a large number of new modelling elements, comprehensive model configurations and tunes planned for release in v3.2.0. 
A comprehensive description of GENIE v3 is much beyond the scope of this article, which will only highlight some of the most recent technical and physics developments, with particular emphasis on new developments that followed the release of v3.0.0\footnote{A complete description of GENIE v3 will appear in  future publications or in the manual available on the GENIE website \url{http://www.genie-mc.org/}}. 

This article is organised as follows: technical developments, with particular emphasis on the core generator framework improvements, a new event library interface that allows experiments to re-use the mature GENIE experimental interfaces with third-party neutrino generators, and the GENIE global analysis of neutrino scattering data are discussed in Sec.~\ref{sec:tech}.
New developments in the description of neutrino, electron and hadron-nucleus scattering, which are discussed in Sec.~\ref{sec:modeling}, are a focal point of this article.
Special emphasis is given in the expanded modelling of the nuclear ground state, the careful validation and improvement of electron-nucleus scattering simulations, new models of zero-pion production (i.e. quasielastic (QE) and multi-nucleon mechanisms such as meson exchange current (MEC) or two-particle-two-hole ($2p2h$) excitations in general),
single-pion production (including both resonant (RES) and non-resonant (NONRES) contributions to the amplitude),
new models for coherent (COH) single-photon production and coherent elastic scattering,
and new advanced models of final state interactions (FSI) delivered through interfaces to the INCL\footnote{ Li\`ege Intranuclear Cascade}~\cite{Mancusi:2014eia} and Geant4~\cite{Wright:2015xia} codes,
a complete new set of high-energy simulation modules including a next-to-leading order (NLO) deep-inelastic scattering (DIS) simulation, and a high-level description of all recent comprehensive configurations and tunes.
Finally, in Sec.~\ref{sec:bsm}, we highlight recent developments in GENIE beyond the Standard Model (BSM) modelling capabilities, which form an important component of the overall program of work in GENIE, in support of the full science program of modern neutrino experiments.

\section{Technical updates}
\label{sec:tech}

One of the most visible updates is the evolution of GENIE into a {\em suite} of separate products, maintained in different repositories. Notable GENIE open-source products include the:
a) {\em Generator}, containing all GENIE physics modules, experimental interfaces (flux and detector geometry drivers) and a host of generic and specialised event generation applications,
b) {\em Reweight}, containing procedures for propagating generator uncertainties,
c) {\em Lamp}, which includes a collection of scripts for building GENIE and necessary external packages,
d) {\em UnitTests}, 
and
e) {\em AVS-CI}, containing GENIE's Automated Validation Suite for Continuous Integration.
Codes for data/MC comparisons that in earlier GENIE v2 revisions existed within the generator, were extracted and formed the basis of additional products that were the focus of substantial development from the core GENIE team over the past few years. They include the:
a) {\em Comparisons}, containing curated archives of neutrino, charged-lepton and hadron scattering data, as well as highly-developed software to produce a comprehensive set of data/MC comparisons, 
b) {\em Prof-GENIE}, implementing the GENIE interface to the Professor tool~\cite{Professor},
and
c) {\em Tuning}, containing the procedures implementing the GENIE global analysis of neutrino scattering data.
The latter group of products plays a central role in the development and characterization of GENIE comprehensive models and tunes and in the GENIE global analysis. While this analysis is in active development, these products do not have open source releases and, therefore, a detailed description of the numerous developments there-in is not in the scope of this brief article. 
All the repositories corresponding to the suite of GENIE products changed from svn~\cite{svn} to git~\cite{git}, and they are hosted in the GENIE organization on GitHub, \url{https://github.com/GENIE-MC}.

At a more detailed level, some of the most visible technical updates were implemented with the goal of supporting an expanded physics content and they range from framework improvements to an interface for a systematic tuning procedure.
Some highlights are listed below.

\subsection{Core software framework improvements}
\label{sec:tech_fmwk}

From the user point of view, the changes in the framework are all related to configuration and mostly geared toward internal consistency. 
As the different code modules are largely independent, it was possible to configure physics parameters to be different for each model leading to the generation of events obtained with inconsistent physics settings.
For example different models could use different values of the coupling constants.
To solve this issue, the concept of \emph{common parameter} was introduced: they are parameters sets that are configurable by the user and yet uniquely defined in memory allocations accessible from every algorithm. 

As it was for version 2, the event generation is subdivided in different processes. 
In this context, processes are labelled after different scattering types (QE, RES, DIS, etc) but they are GENIE terminology to identify different event generation algorithms that are tailored for specific final states.
In general, processes are not universal and their definitions are only valid within the GENIE software.
This modular structure has a large degree of configuration: for each process, the system offers a number of alternative models to be used for event generation, see Sec.~\ref{sec:modeling}. 
In previous GENIE releases, only one model-process mapping was suggested by the out-of-the-box configuration, despite the availability of alternative models.
Yet, there was no guidance on how to correctly use different configurations according to author and developers.
Hence, it was easy to come up with inconsistencies between the model configuration for different processes that were supposed to be used together to get a correct comprehensive physics simulation. 
An example of inconsistent configuration is using the Valencia model (see Sec.~\ref{sec:cc0pi}) with a relativistic Fermi gas nuclear model. 
This issue was addressed in GENIE v3 by introducing the concept of comprehensive model configuration (CMC) that is a consistent process-model association.
Considering that GENIE already has about 20 different processes for neutrinos alone, CMC definitions are quite complex objects and they need to be effectively named so that the community can use them unambiguously. 
For this purpose, the collaboration developed a specific naming convention that is described in the manual. 
Section~\ref{sec:MEnu_tunes} will summarise the names and the physics relevant for for this paper. 
These CMCs are aimed at specific types of experiment and can be expanded according the needs of collaborations.

A final element to be mentioned is related to the internal PDG~\cite{PDG} library. 
The PDG values used so far in GENIE have not been changed to keep the predictions consistent with past versions. 
Now users will have the possibility to use different PDG library configurations that are tune dependent.

\subsection{The new GENIE tuning process}
\label{sec:tech_tuning}
Tuning is a necessary step for all MC generators. 
In the specific case of neutrino generators, it is required to merge together different models in order to avoid double counting, since there is no single model able to cover all possible interactions across the whole energy range. 
Indeed, 
development of a global analysis of scattering data for the tuning and uncertainty characterization of comprehensive neutrino interaction models 
has been a central activity of the core GENIE team over the past few years.
The GENIE Generator is the main outlet for the GENIE global analysis results, and our goal is that, for each supported comprehensive model, several selected tuned versions shall be made available. 

The GENIE global analysis was made possible through the continued development of curated data archives, and their successful interface to the Professor tool~\cite{Professor}.
This interface enabled the efficient implementation of complex multi-parameter brute-force scans and removed substantial global analysis limitations by decoupling it from event reweighting procedures that, for all but the most trivial aspects of our physics domain, require substantial development time and are not exact, or even possible at all.
Specifically, Professor `reduces the exponentially expensive process of brute-force tuning to a scaling closer to a power law in the number of parameters, while allowing for massive parallelisation'~\cite{ProfessorWebPage}.

This concept goes beyond the existing reweighting scheme since it allows the tuning of parameters that are not normally reweightable.
We expect to be able to develop a reweighting tool based on this method for GENIE v4.  After a specific experimental flux is defined as an input, the phase space of each interaction can be decomposed in bins seen as an observable. 
Eventually we expect the users and experiments to build their own response functions to allow the reweighting of their predictions according to the statistical output of the tunes based on this technology.

\subsection{Event library}
\label{sec:tech_interfaces}
The neutrino flux and geometry interfaces and the wealth of mature and well-tested drivers implementing these interfaces constitute one of the most well-known and desirable GENIE features that has catalysed GENIE adoption and enabled seamless integration in the full MC simulation chain of all current and near-future neutrino experiments. No other physics generator provides an equivalent and equally comprehensive and mature toolkit and experimental interfaces. There is a strong community desire to reuse the GENIE experimental interfaces to test alternative physics generators. This drove the implementation of an ``Event Library'' interface~\cite{EventLibrary} and the development of a generic {\em EvtLib} GENIE generator.

The purpose of the {\em EvtLib} generator is to read from an external library of cross sections and pre-computed final particle kinematics (most likely computed using an alternative neutrino generator). 
For each interacted neutrino selected by GENIE, the generator will use the appropriate cross section from the file, and then use the kinematics from the library entry with the closest-matching energy. 
Within the limits of the library statistics, this will then reproduce the physics of the  external generator, but making use of the flux and geometry handling of GENIE.
The details of the event library file structure are described in the code and in the manual~\cite{manual}.


\section{Interaction modeling improvements}
\label{sec:modeling}

Neutrino-nucleus interactions are very important to many experiments and this remains a central area of effort within the GENIE collaboration.  
Charged-current neutrino interactions without final-state mesons (CC0$\pi$ interactions) will dominate the expected signal in future precision oscillation measurements by the Short-Baseline Neutrino program \cite{SBNPaper} and Hyper-Kamiokande \cite{HyperK}.  
Significant GENIE development effort has recently been devoted to the implementation of new models of quasielastic and $2p2h$ interactions (see Sec.~\ref{sec:cc0pi}). 
Many channels will be important for the upcoming Deep Underground Neutrino Experiment (DUNE) \cite{DUNETDRVol1}, especially because the higher average beam energy will enhance the role of more inelastic event topologies. 
Resonance production and FSI will be very important for DUNE. 
Recent improvements in those areas are discussed in Secs.~\ref{sec:PiProd} and \ref{sec:MEnu_fsi}, respectively.

At the same time, significant efforts have gone into new capabilities at very low energies (Sec.~\ref{sec:cevns}) and very high energies (Sec.~\ref{sec:hedis}). 
In addition, the importance of electron scattering~\ref{sec:emode} to determine nuclear structure~\ref{sec:ground_state} and vector interactions is expanding. 
These three directions greatly enhance the reach of GENIE into new experiments. 

\subsection{Nuclear ground state}
\label{sec:ground_state}

At energies relevant for accelerator neutrino experiments, a variety of nucleon-level hard scattering processes (principal interactions), such as resonance production, must be considered when preparing a comprehensive lepton-nucleus interaction model for use in an event generator. However, two aspects of such a comprehensive model will be common to all interaction modes: a description of the nuclear ground state (the subject of this section) and a treatment of intranuclear rescattering due to hadronic final-state interactions (the subject of Sec.~\ref{sec:MEnu_fsi}).

There are a variety of methods to model the nuclear ground state. 
Improving these models is an ongoing process within GENIE.
At present, the nuclear ground state is represented by a spectral function $P(\mathbf{p}, E)$ which describes the probability that a nucleon involved in a lepton-nucleus interaction will have an initial 3-momentum $\mathbf{p}$ and removal energy $E$. 
In the historical default model used since GENIE v2, the initial nucleon momentum is sampled according to the relativistic Fermi gas (RFG) treatment of Bodek and Ritchie \cite{BodekRitchieRFG}. 
This version of the RFG has non-interacting nucleons up to the Fermi momentum $k_F$, which is determined from inclusive electron scattering.
It also accounts for short-range nucleon-nucleon correlations \cite{SRC} by adding a high-momentum tail above $k_F$ to the usual distribution. 
A fixed, isotope-specific removal energy $E$ is used in all cases. 
Pauli blocking in quasielastic interactions is implemented by requiring the final-state nucleon momentum to exceed $k_F$.  
The Bodek-Ritchie RFG continues to be used in multiple GENIE CMCs mainly for higher energy processes and connection with past modeling.

Two newer nuclear model implementations are available for all target nuclides in GENIE v3.2. 
The first of these is a local Fermi gas (LFG) model based on the work of the Valencia group~\cite{ValenciaModel} and various other publications. 
Under this approach, the high-momentum tail is neglected, and $k_F$ is a function of radius obtained by Fourier transforming the nucleon number density $\rho(r)$.
The implementation of this model underpins a related treatment of quasielastic and two-particle-two-hole interactions (see Sec.~\ref{sec:cc0pi}), in which nuclear effects such as long-range correlations and Coulomb corrections are handled according to the same local density approximation in the Valencia model.

\begin{figure}[htp]
\centering
\includegraphics[width=0.53\textwidth]{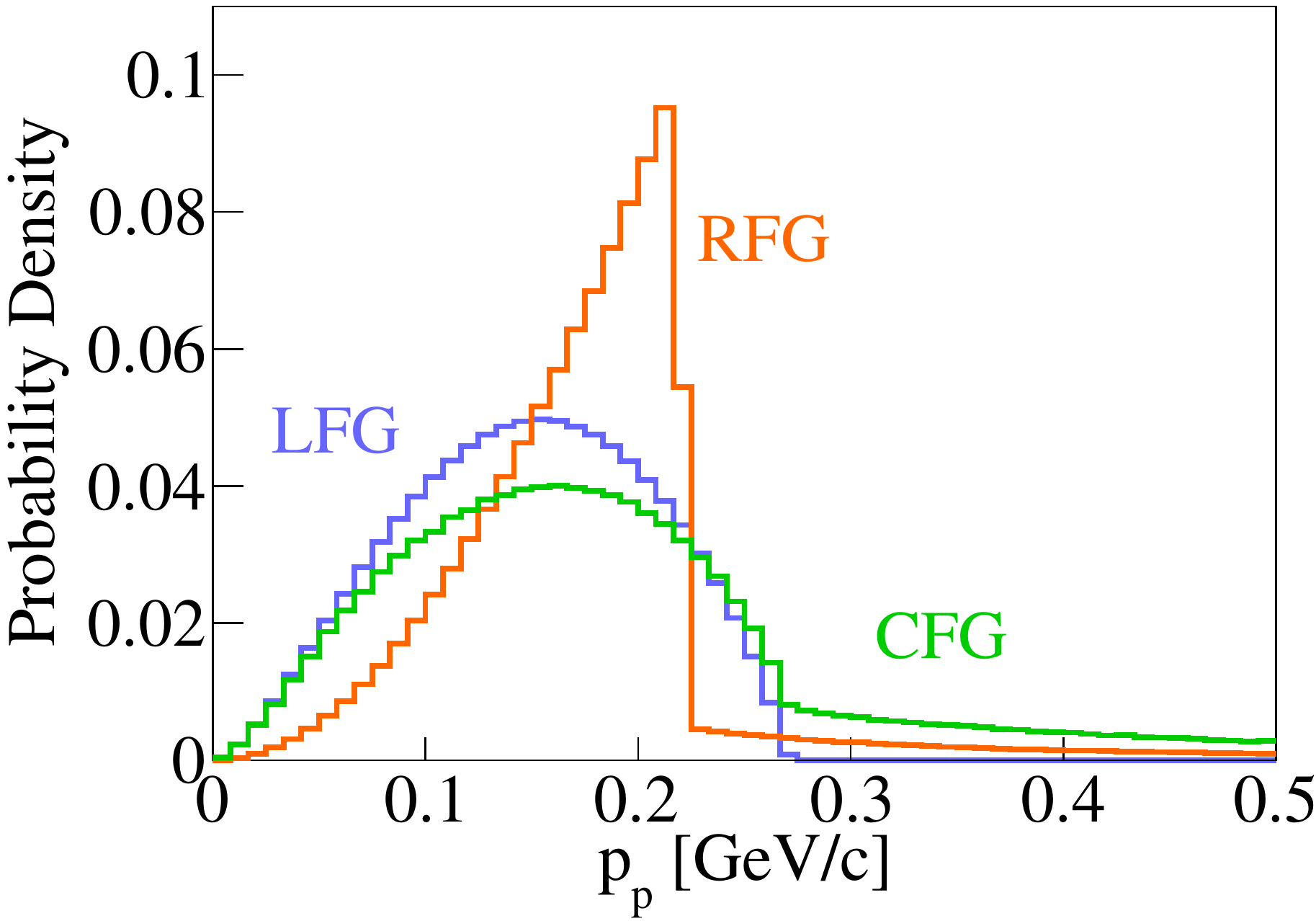}
\caption{Initial nucleon momentum magnitude distributions according to the GENIE implementation of relativistic Fermi gas, local Fermi gas and correlated Fermi gas models.}
\label{fig:GCF}
\end{figure}

A variation of the original LFG model, called the correlated Fermi gas (CFG)~\cite{SRCFractionPaper}, has also been added in GENIE v3.2. 
The CFG keeps the radial dependence of the LFG model while adding a high-momentum tail that lies above the local $k_F$. 
The original LFG distribution is renormalized to ensure that a given fraction of initial-state nucleons is found in the tail. 
The current default of 20\% is based on electron scattering measurements \cite{SRCFractionPaper} and may be adjusted in future GENIE tuning efforts. 
Both the LFG and CFG implementations in GENIE use a fixed nucleon removal energy which is identical to that used by the Bodek-Ritchie RFG. 
Fig.~\ref{fig:GCF} shows the  $|\mathbf{p}|$ distribution predicted by each of the three models of the nuclear ground state discussed above, where $\mathbf{p}$ is the initial nucleon momentum.

\subsection{Electron-nucleus scattering}
\label{sec:emode}

Since neutrinos and electrons are both leptons, they interact with atomic nuclei in similar ways. Electrons interact via a vector current 
and neutrinos interact via vector and axial-vector 
currents.
Electron-nucleus scattering data allow for more precise measurements than $\nu$-nucleus scattering due to an interaction rate that is $O(10^7)$ times higher, thanks to the high cross section.
Furthermore, the knowledge of the incoming flux is more constrained.
Specifically, monochromatic beams allow for proper kinematic reconstruction, reducing the systematic uncertainty.
These higher quality data allow for powerful model constraints on the vector part of lepton-nucleus cross sections. 
Key nuclear effects common to electron and neutrino probes, such as the nuclear ground state and hadronic FSIs, can also be studied in detail. 

In recent versions of GENIE, significant improvements were made for aspects of both neutrino and electron cross section modeling~\cite{papadopoulou2020inclusive}. 
Significant errors were corrected, including a mistake in the mathematical expression used for the QE differential cross section, a missing Lorentz boost in the $2p2h$ interaction affecting both electrons and neutrinos, and incorrect electron couplings used in the RES interactions.  
Wherever possible, the electron treatment was updated to be significantly more similar to the neutrino one and to use the same computer code.

\begin{figure}[htp]
\centering
\includegraphics[width=0.55\textwidth]{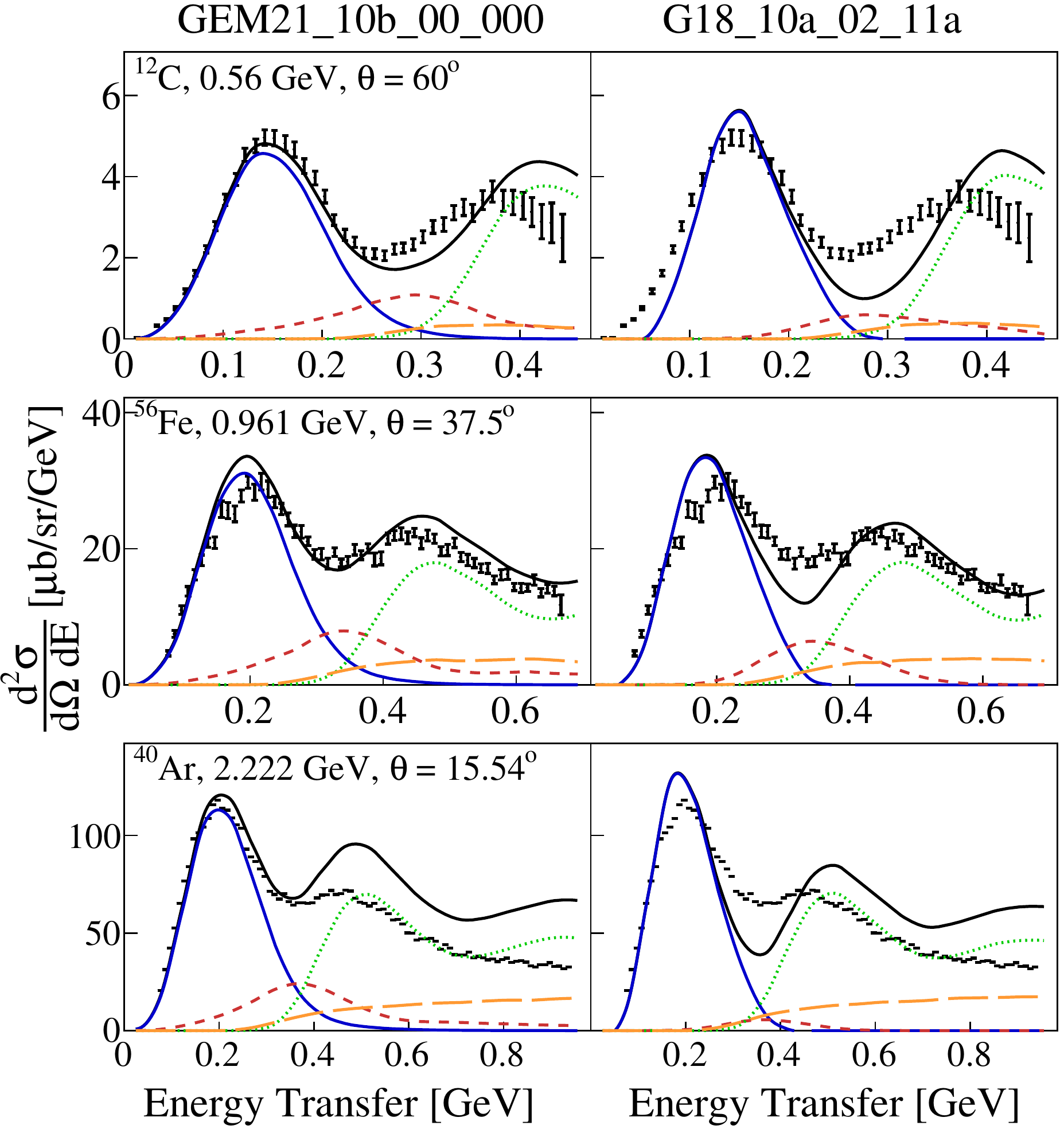}
\caption{Comparison of inclusive $(e,e')$ scattering cross sections
  for data and for GENIE.  (left) data vs \texttt{GEM21\_10b\_00\_000} and (right) data vs
  \texttt{G18\_10a\_02\_11a}.  (top) carbon at $E_0=0.56$ GeV, $\theta_e = 60^\circ$ and
  $Q^2_{\text{QE}}\approx 0.24$ GeV$^2$, (middle)
  iron at $E_0=1$ GeV, $\theta_e = 37.5^\circ$ and $Q^2_{\text{QE}}\approx 0.32$
  GeV$^2$, and (bottom) argon at $E_0=2.22$ GeV,
  $\theta_e = 15.5^\circ$ and $Q^2_{\text{QE}}\approx 0.33$
  GeV$^2$. Black points show the data, solid black
  lines show the total GENIE prediction, colored lines show the
  contribution of the different reaction mechanisms: (blue) QE,
  (red) MEC, (green) RES and (orange) DIS.}
\label{fig:eCIncluisve}
\end{figure}

The GENIE collaboration is in the process of benchmarking the electron scattering predictions against existing inclusive electron scattering data for different target nuclei, beam energies and scattering angles~\cite{papadopoulou2020inclusive}, as can be seen in \figurename~\ref{fig:eCIncluisve}.
The physics content of the model configurations shown is described in Sec.~\ref{sec:MEnu_tunes}\footnote{Please note that the GENIE configuration referred to in Ref.~\cite{papadopoulou2020inclusive} as GSuSAv2 or \texttt{GTEST19\_10b\_00\_000} is now labeled as \texttt{GEM21\_11b\_00\_000} in GENIE version 3.02.00. The previous ``test'' configuration was promoted to a full CMC.}. 
The agreement is very good for the kinematic region dominated by QE processes due to the bug fixes and adoption of newer models.  On the other hand, the simulation is well above the data in the resonance region. This is largely due to deficiencies in modelling of the fundamental scattering process rather than the treatment of nuclear effects. In particular, the existing GENIE tunes to measurements with hydrogen and deuterium targets~\cite{free_nucleon_tune} only use neutrino (as opposed to electron) data. As a partial accounting, we have added the Bosted-Christy model~\cite{Bosted:2007xd,Christy:2007ve} as an alternate cross section for use in user-driven reweighting.

The community is also in the process of reviewing and improving the electron scattering data that will be useful for our validation purposes. 
For example, high-statistics datasets from the CLAS6 detector at Jefferson Laboratory were analysed ~\cite{MarianaThesise4nu} 
and new  experiments ($e4\nu$, Mainz, and LDMX) designed to support  neutrino interaction experiments will start data taking in 2021~\cite{e4nuProposal,MaintzProposal,akesson2018light}.  
Major emphasis in these new measurements will be placed on a detailed description of the hadronic part of the final state.


\subsection{ CC0$\pi$ cross sections }
\label{sec:cc0pi}
Initial efforts to isolate CCQE interactions in neutrino-nucleus scattering data proved to have significant model dependence.  
Hence, experimental attention is now focused on CC0$\pi$ event topologies which involve three underlying processes within GENIE. 
CCQE and $2p2h$ are the main contributors; nonetheless, pion production followed by intranuclear absorption also contributes significantly to this channel.

The GENIE v2 historical default model for charged-current quasielastic scattering was based on the Llewellyn Smith \cite{LSQE} formalism: the expression for the hadronic part of the cross section is taken to be the same as for a free nucleon. The corresponding cross section for scattering on a complex nucleus is then computed by correcting for Pauli blocking and binding energy, averaging over the initial nucleon momentum distribution (see Sec.~\ref{sec:ground_state}), and then multiplying by the total number of neutrons (protons) for an incident neutrino (antineutrino).

The Llewellyn Smith approach is still available in GENIE v3.2 and remains a good model for neutrino energies above roughly 2 GeV. Two additional CCQE models which contain details that are important for lower energy neutrinos have now also been implemented. 
One of these is based on the formalism of the Valencia group \cite{ValenciaModel} and makes two major refinements beyond Llewellyn Smith. First, long-range nucleon correlations are treated in a Random Phase Approximation (RPA) approach.
Corrections for these are included as density-dependent modifications to the free-nucleon hadronic tensor. Second, corrections for the final-state Coulomb interaction of the outgoing charged lepton are introduced using a strategy similar to the ``modified effective momentum approximation'' proposed by Engel \cite{EngelCoulomb}. 
The radial dependence of both of these nuclear effects is taken into account by relying on the local Fermi gas model described in Sec.~\ref{sec:ground_state}.


In the GENIE implementation of the Valencia model, a correction for binding energy is made by assigning an off-shell total energy to the initial struck nucleon. The QE cross section is then calculated under the de Forest prescription \cite{DeForest1983}: an effective energy transfer is used which is reduced by the amount of energy needed to put the initial nucleon on the mass shell. 
Nieves \textit{et al.} also recommend using an effective energy transfer in the original Valencia model publication, but their approach is different. Rather than considering nucleon knock-out, the authors reduce the energy transfer based on the difference in ground-state masses between the initial nucleus and a final nucleus that includes the outgoing nucleon\footnote{A correction is also made for the difference in Fermi energies between the final and initial nucleon species, see Eq.~(43) in Ref.~\cite{ValenciaModel}.}.

\begin{figure}
    \centering
    \includegraphics[width=0.49\textwidth]{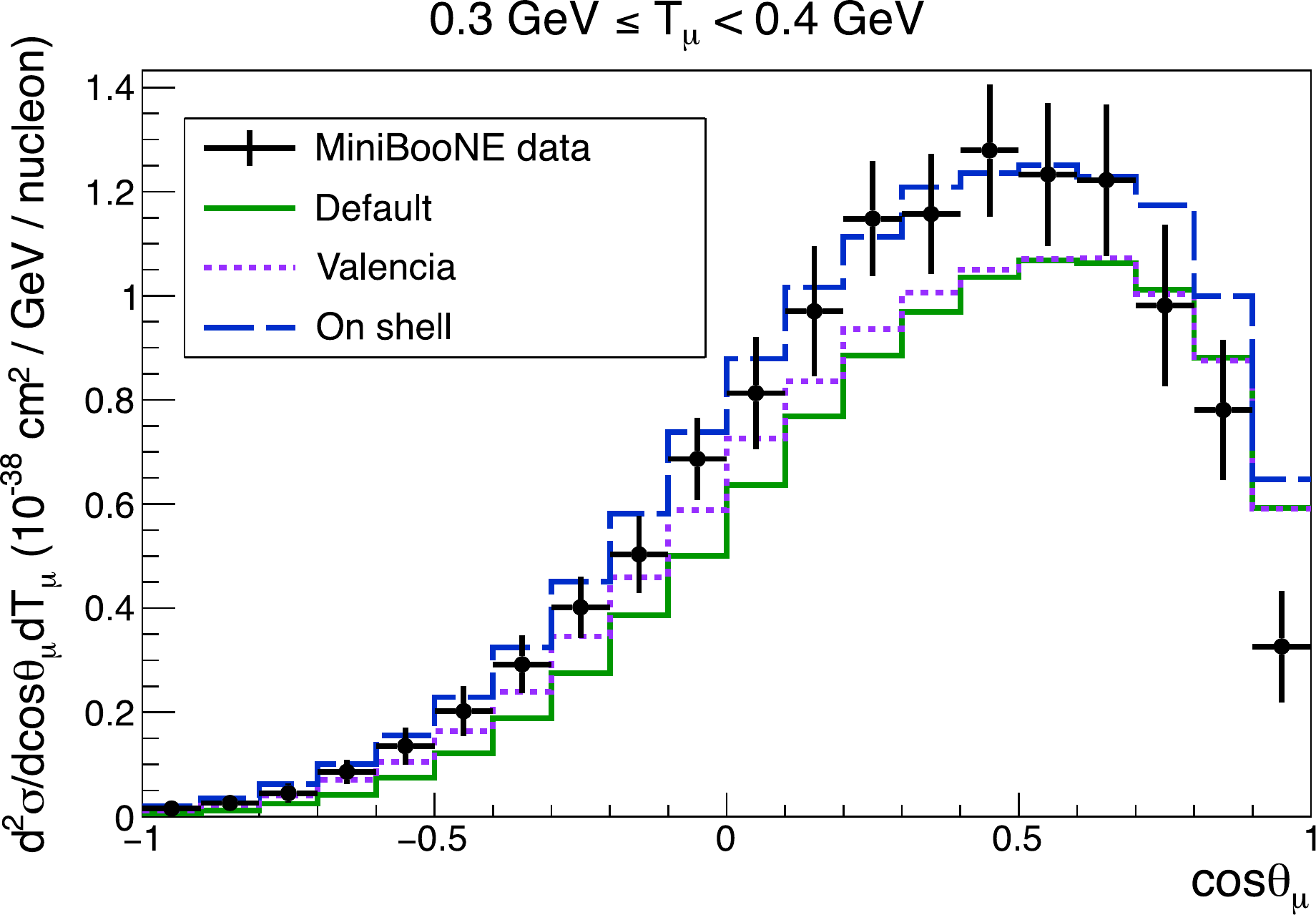}
    \hfill
    \includegraphics[width=0.49\textwidth]{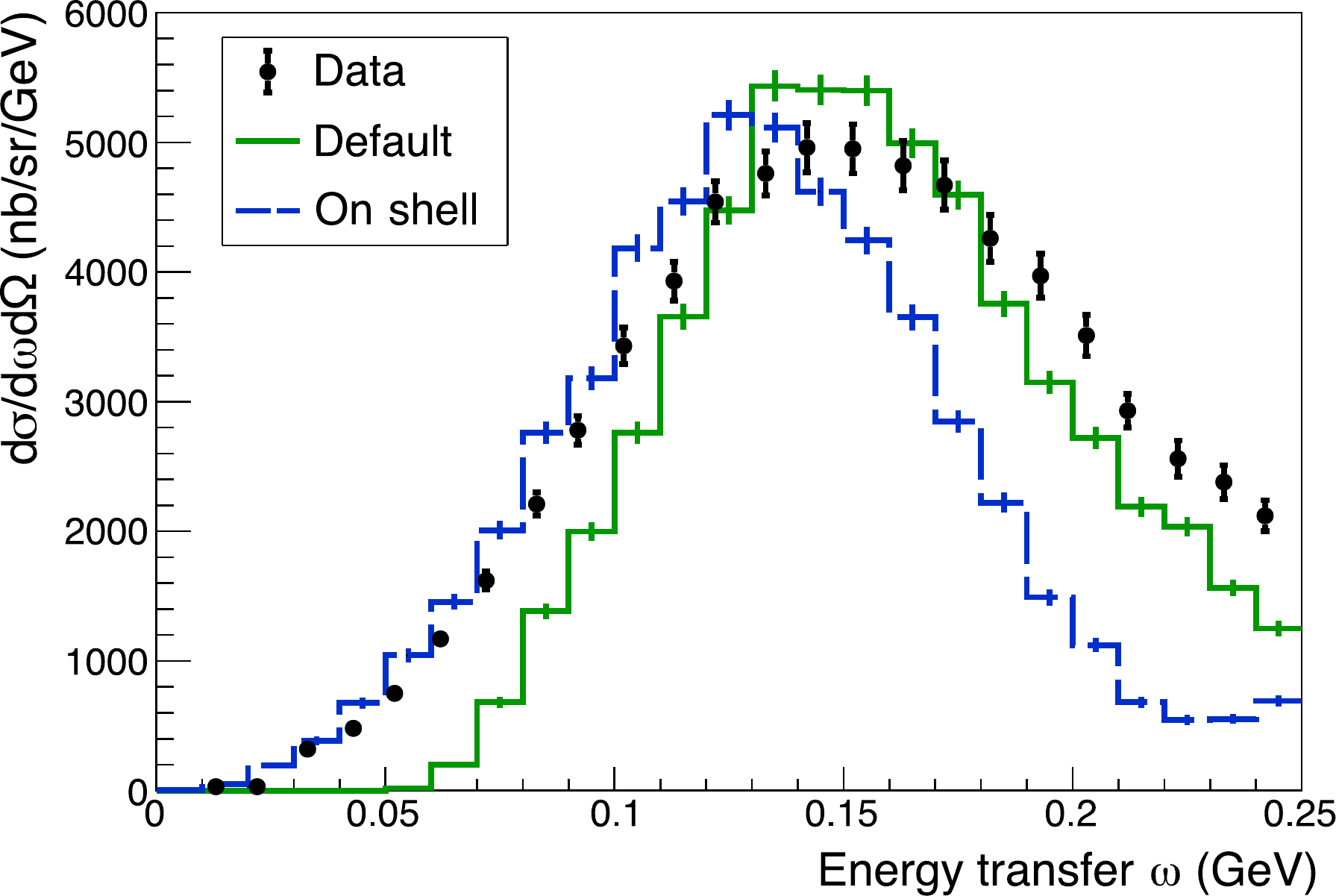}
    \caption{Impact of nucleon binding energy effects on GENIE cross section predictions. LEFT: MiniBooNE double-differential CC0$\pi$ cross section compared to GENIE \texttt{G18\_10a\_02\_11b} predictions calculated with variations to the binding energy correction. RIGHT: Differential cross section predictions for inclusive 560 MeV electron scattering on \isotope[12]{C} at $\theta = 60^\circ$. The solid green histogram shows the nominal prediction using the \texttt{G18\_10a\_02\_11b} CMC. The dashed blue histogram shows the same cross section when no correction for binding energy is applied. Data points are taken from Ref.~\cite{Barreau1983ht}.} 
    \label{fig:BE_plot}
\end{figure}

The left-hand plot in Fig.~\ref{fig:BE_plot} shows a comparison of GENIE calculations to MiniBooNE CC0$\pi$ data~\cite{MiniBooNECC0pi} in which both the default binding energy treatment (solid green) and original Valencia model approach (dotted violet) have been applied. Noticeably better agreement is achieved in the backwards-angle bins with the Valencia procedure. Interestingly, further overall improvement is seen if binding energy corrections are neglected entirely (dashed blue).
The importance of binding energy corrections for low-energy lepton-nucleus scattering is further highlighted in the right-hand plot of Fig.~\ref{fig:BE_plot}, which shows a comparison of GENIE predictions to data obtained for 560 MeV electrons on \isotope[12]{C} at a scattering angle of $\theta = 60^{\circ}$. Here the default GENIE approach to binding energy (solid green) achieves much better agreement with the measured QE peak location than the prediction where binding energy is neglected entirely (dashed blue).
The $Q^2$ values represented by the data points in the right-hand plot (ranging between 0.17--0.31 GeV$^2$) are roughly comparable to the MiniBooNE flux-averaged mean $Q^2 \approx 0.4$ GeV$^2$ predicted by the GENIE \texttt{G18\_10a\_02\_11b} simulation.


A second new CCQE model in GENIE \cite{GSuSAv2} implements the SuSAv2 treatment \cite{SuSAv2QE}. Under this approach, the nuclear responses are calculated using scaling functions based on Relativistic Mean Field (RMF) theory. A precomputed table of these responses, defined on a two-dimensional grid in energy and momentum transfer, is interpolated for efficient sampling of final-state lepton kinematics. Handling of such tabular input for QE and $2p2h$ models is an important new capability added in v3.2.
A factorisation strategy is employed to simulate the outgoing nucleon: the leptonic 4-momentum transfer is applied to a nucleon drawn at random from the initial-state single nucleon distribution (see Sec.~\ref{sec:ground_state}). The limitations of this approximation are considered in Ref.~\cite{GSuSAv2}.

New models of $2p2h$ interactions have also been recently implemented in GENIE following the Valencia \cite{Nieves,Gran:2013kda} and SuSAv2 \cite{RuizSimo2016,Simo2017} approaches. These provide theory-driven alternatives to the Empirical model \cite{Katori2013} available since late releases of GENIE v2. In contrast to the QE case, the SuSAv2-MEC model is based on a relativistic Fermi gas description of the nucleus. Both new $2p2h$ models rely on an implementation strategy similar to the one used for SuSAv2 QE: inclusive differential cross sections are calculated using tables of nuclear responses \cite{Schwehr2017}, and the sampled 4-momentum transfer is then imparted to a cluster of two nucleons chosen from the single-nucleon ground-state nuclear model. Separate nuclear response tables are provided based on the isospin composition of the struck nucleon pair ($nn$, $pn$, or $pp$), which is chosen by comparing the relative contributions to the inclusive differential cross section at fixed lepton kinematics.
The combined Valencia QE+$2p2h$ model is available in GENIE only for CC neutrino scattering, while SuSAv2 may be applied to electron scattering as well.

\begin{figure}
    \centering
    \includegraphics[width=0.49\textwidth]{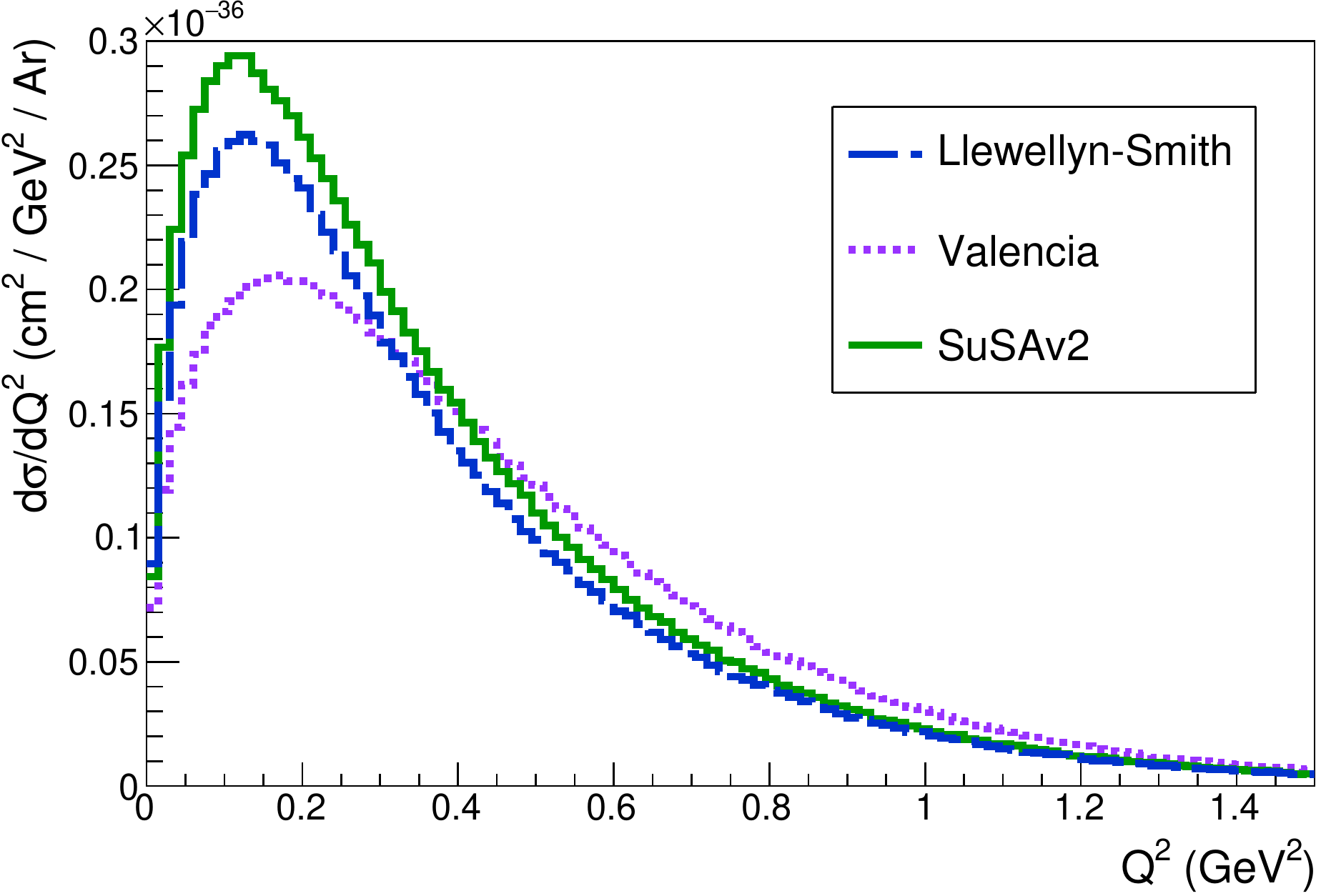}
    \hfill
    \includegraphics[width=0.49\textwidth]{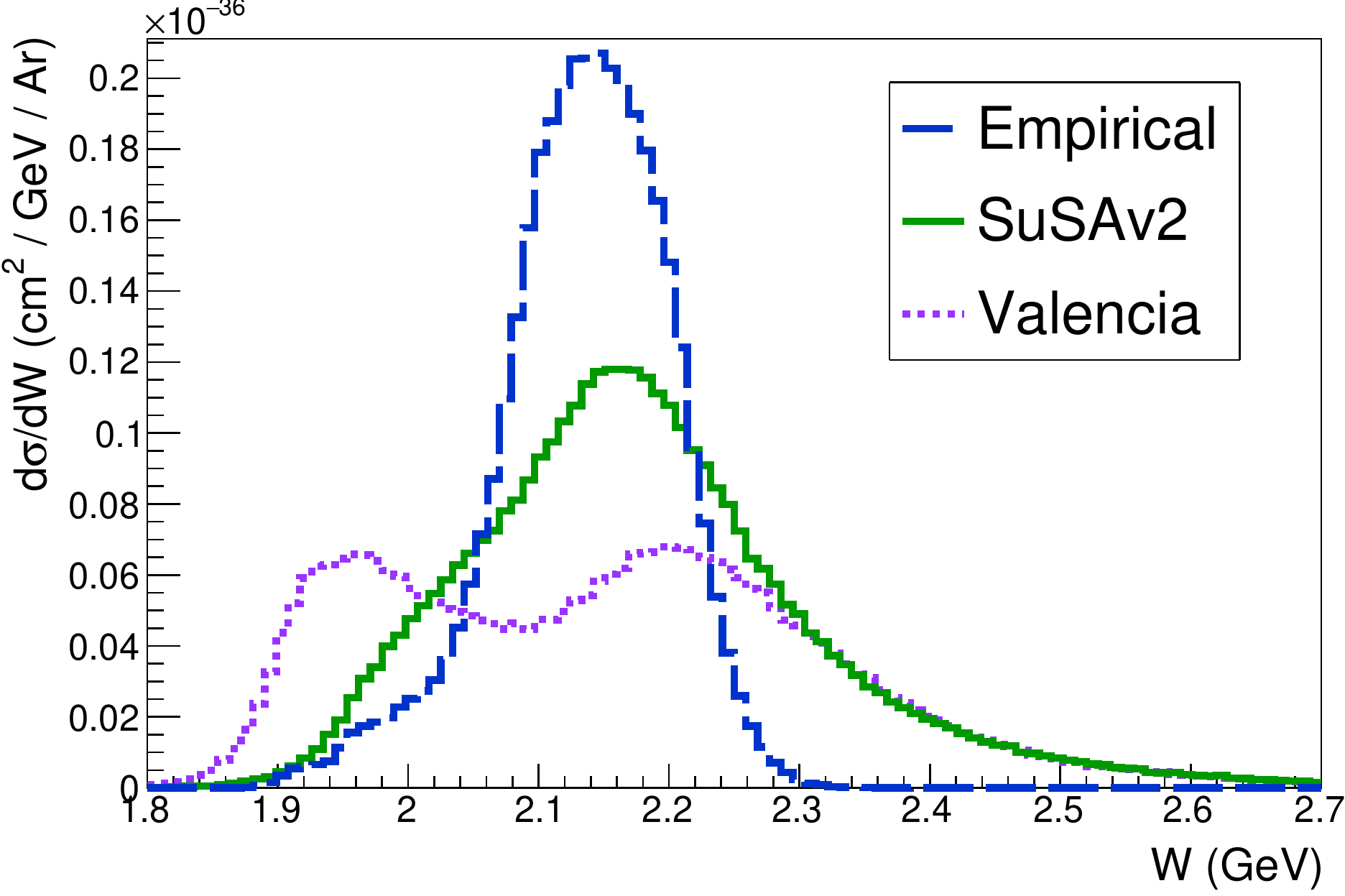}    
    \caption{Differential cross sections predicted by several different GENIE treatments of $\nu_\mu$ charged-current quasielastic (left) and two-particle-two-hole (right) interactions on argon. All distributions shown are averaged over the MicroBooNE $\nu_\mu$ flux.}
    \label{fig:CC0pi_truth_plots}
\end{figure}

The left (right) plot in Fig.~\ref{fig:CC0pi_truth_plots} illustrates some representative differences between the three GENIE CCQE ($2p2h$) models described above. The RPA corrections included in the Valencia CCQE model lead to a suppression of low-$Q^2$ events, shown here for $\nu_\mu$ scattering on argon in MicroBooNE. All three $2p2h$ models predict distinct distributions of the hadronic invariant mass $W$, with the Valencia calculation uniquely splitting the strength into two peaks. Neutrino detectors capable of measuring pairs of final-state nucleons, such as liquid argon time projection chambers \cite{ArgoNeuTPairs}, may provide helpful constraints on these $2p2h$ model differences in the future.


\subsection{ Pion production }
\label{sec:PiProd}

Treatment of pion production in GENIE is of great interest because many aspects of the underlying theory are complicated and poorly understood.  
The existing GENIE models are all based on the phenomenological approach of  Rein, and Sehgal (RS model) \cite{Ravndal:1973xx,Rein:1980wg}, that  aimed at describing pion production in the resonance region using nucleon-to-resonance transition matrix elements calculated with the relativistic quark model of Feynman-Kislinger-Ravndal. The original RS model (without the interference between resonances) has always been in GENIE. The non-resonant background (NRB) in GENIE is simulated by DIS contribution~\cite{Paschos:2001np} with the structure functions proposed by Bodek and Yang \cite{Bodek:2004pc,Bodek:2010zz}. 
This model extends down to $\pi$N threshold; its normalization is adjusted in the ``resonance-dominated'' region so that the summed response in this region agrees with $\nu$\,H/D inclusive cross section data.  
In v2.10, improvements by by Kuzmin, Lyubushkin, and Naumov (KLN) \cite{Kuzmin:2004ya} and by Berger and Sehgal (BS) \cite{Berger:2007rq} were introduced to account for nonzero lepton mass, lepton polarization, and pion pole contributions. 
At the same time, updated form factors for pion production were proposed by Graczyk and Sobczyk \cite{Graczyk:2007bc} and used by the MiniBooNE Collaboration \cite{Nowak:2009se}. 
 The RS model parametrises the axial transition form factors in terms of a  common parameter, the axial-vector mass, $M_A^{\text{RES}}$, which is adjusted in each physical tune, and the default value is 1.12~GeV \cite{Kuzmin:2006dh}. 
 Details on the previous implementations are given in Refs.~ \cite{manual,Alam:2015nkk}, and the approach to combining resonant and non-resonant contributions taken in GENIE 3.0 is described in 
Ref.~\cite{free_nucleon_tune}.  
We discuss improvements to the existing models and the new model in GENIE 3.2 here. 
In the current version, normalization of the Breit-Wigner distributions can be optionally switched on/off. 
In addition, bugs were fixed and the code was optimized.

 This GENIE release includes an implementation of the new single-pion ($1\pi$) production model \cite{Kabirnezhad:2017jmf,Kabirnezhad:2017xzx1,Kabirnezhad:2016nwu,Kabirnezhad:2017dui} (the MK model) based on the Rein's formalism~\cite{Rein:1987cb} and the KLN/BS lepton-mass treatment of Refs~\cite{Kuzmin:2006dh, Berger:2007rq}.
 The MK model includes several significant extensions from the current implementations of the $1\pi$ production models in GENIE, especially with the proper accounting for interference between resonances and NRB. 
 Rather than taking a strictly empirical approach, the NRB contribution is provided by generalized Born graphs for the $1\pi$ production based on a chiral SU(2) non-linear $\sigma$ model according to the Hern\'andez, Nieves, and Valverde (HNV) approach; vector form factors are also updated.  
 The MK model is the first resonance model implementation in GENIE that fully incorporates  the interference effects and predictions for pion angular distributions.  

 \begin{figure}
 \centering
 \includegraphics[width=\textwidth]{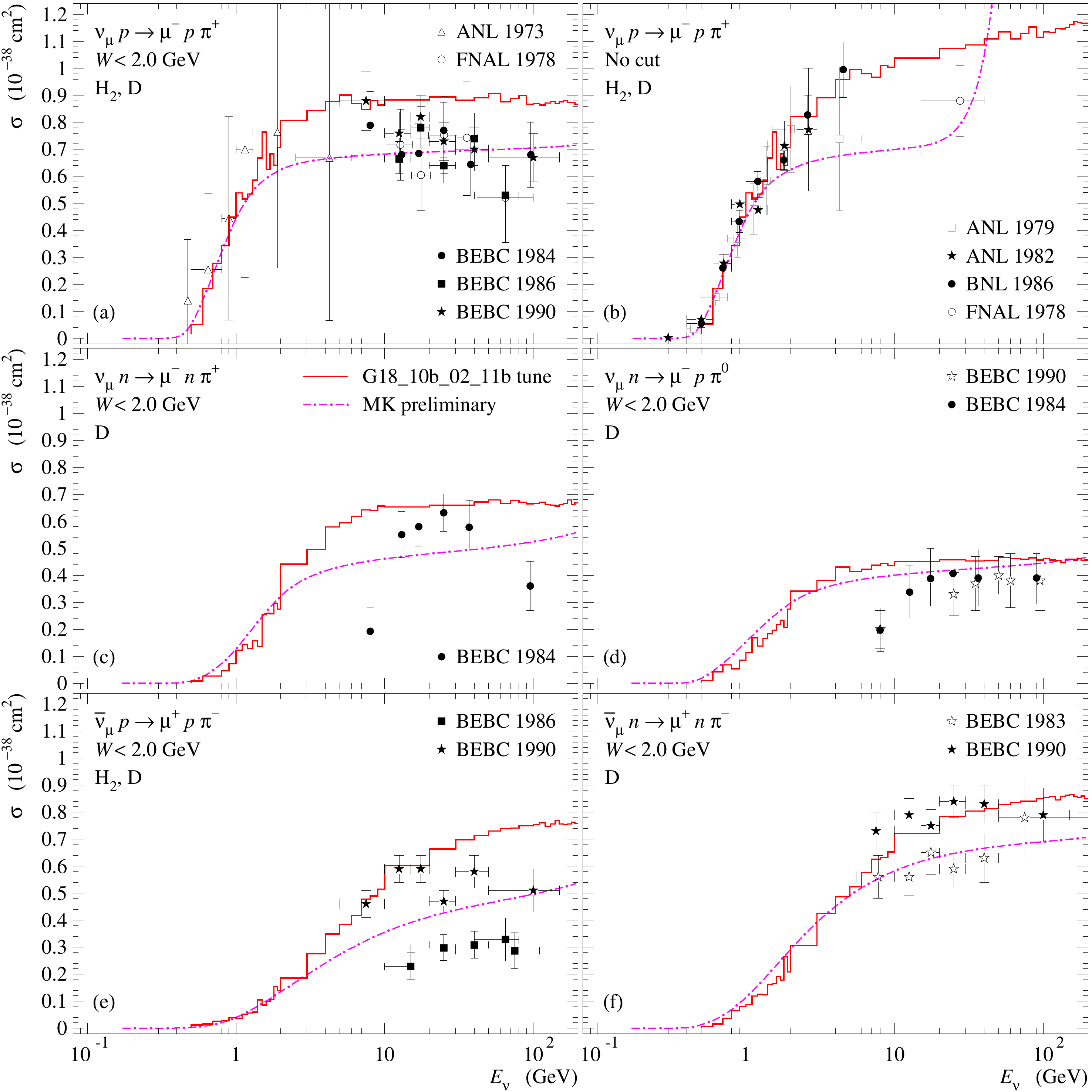}
 \caption{Total cross sections of the reactions
          $\nu_\mu p\to\mu^-p\pi^+$ (a), (b),
          $\nu_\mu n\to\mu^-n\pi^+$ (c),
          $\nu_\mu n\to\mu^-n\pi^+$ (c),
          $\nu_\mu n\to\mu^-p\pi^0$ (d),
          $\overline{\nu}_\mu p\to\mu^+p\pi^-$ (e), and
          $\overline{\nu}_\mu n\to\mu^+n\pi^-$ (f)
          measured at
          ANL 1973 \cite{Campbell:1973wg},
              1979 \cite{Barish:1978pj},
              1982 \cite{Radecky:1981fn},
          BNL 1986 \cite{Kitagaki:1986ct},
          FNAL 1978 \cite{Bell:1978rb,Bell:1978qu}
          CERN BEBC 1983 \cite{Allasia:1983qh},
                    1984 \cite{Barlag:1984uga},
                    1986 \cite{Allen:1985ti}, and
                    1990 \cite{Allasia:1990uy}
          with cut of $W < 2$ GeV and with no cut on $W$,
          in comparison with 
          \texttt{G18\_10b\_02\_11b} tune and
          MK model (preliminary) incorporated in GENIE~3.
          The error bars are the statistical and systematic errors added quadratically.
         }
 \label{Fig:tot1pixsec}
 \end{figure}

 Figure \ref{Fig:tot1pixsec} shows the total CC$1\pi$ production cross sections
 for $\nu_\mu$ and $\overline{\nu}_\mu$ induced reactions
 with the bare nucleons as predicted using the \texttt{G18\_10b\_02\_11b} tune
  and MK model \cite{Kabirnezhad:2017jmf,Kabirnezhad:2017xzx1,Kabirnezhad:2016nwu,Kabirnezhad:2017dui},
 in comparison with the experimental data 
 using bubble-chambers filled with the hydrogen and deuterium.
 Data from CERN BEBC \cite{Allasia:1983qh,Barlag:1984uga,Allen:1985ti,Allasia:1990uy} are used without modification.
 The data of ANL 1982 \cite{Radecky:1981fn} and BNL 1986 \cite{Kitagaki:1986ct}
 are reanalysed in Ref. \cite{Rodrigues:2016xjj}.
 The data of CERN BEBC 1990 \cite{Allasia:1990uy}
 with the cut $W < 2$ GeV are revised in Ref. \cite{Hawker:02}.
 The data of ANL 1973 \cite{Campbell:1973wg} and FNAL 1978 \cite{Bell:1978rb,Bell:1978qu}
 with the cut $W < 2$ GeV are obtained as the cross sections of $\Delta$ production.

\subsection{ COH Gamma }

Neutral-current photon emission reactions with nucleons and nuclei are important backgrounds for $\nu_\mu \to \nu_e$ ($\overline\nu_e \to \overline\nu_e$) appearance oscillation experiments where electromagnetic showers instigated by electrons (positrons) and photons are hard to distinguish. 
For example, it has implications for the T2K oscillation analyses \cite{Wang:2015ivq, Abe:2017vif}. 
In the few-GeV region, these reactions are dominated by the weak excitation of the $\Delta(1232)$ resonance and its subsequent decay into $N$ and $\gamma$ and this process has been available in GENIE since version~2.
The coherent reaction channel (COH gamma), where the nucleus returns to its ground state after emitting a gamma ray, has a small ($5$ to $50$ times smaller than incoherent photon emission, depending on the neutrino energy) but sizable contribution particularly in the forward direction. 
For this reason, it is a background for some of the BSM candidates to explain the MiniBooNE anomaly \cite{Machado:2019oxb} (see also Sec. \ref{sec:bsm_darknu}). 

\begin{figure}
    \centering
    \includegraphics[width=.49\textwidth]{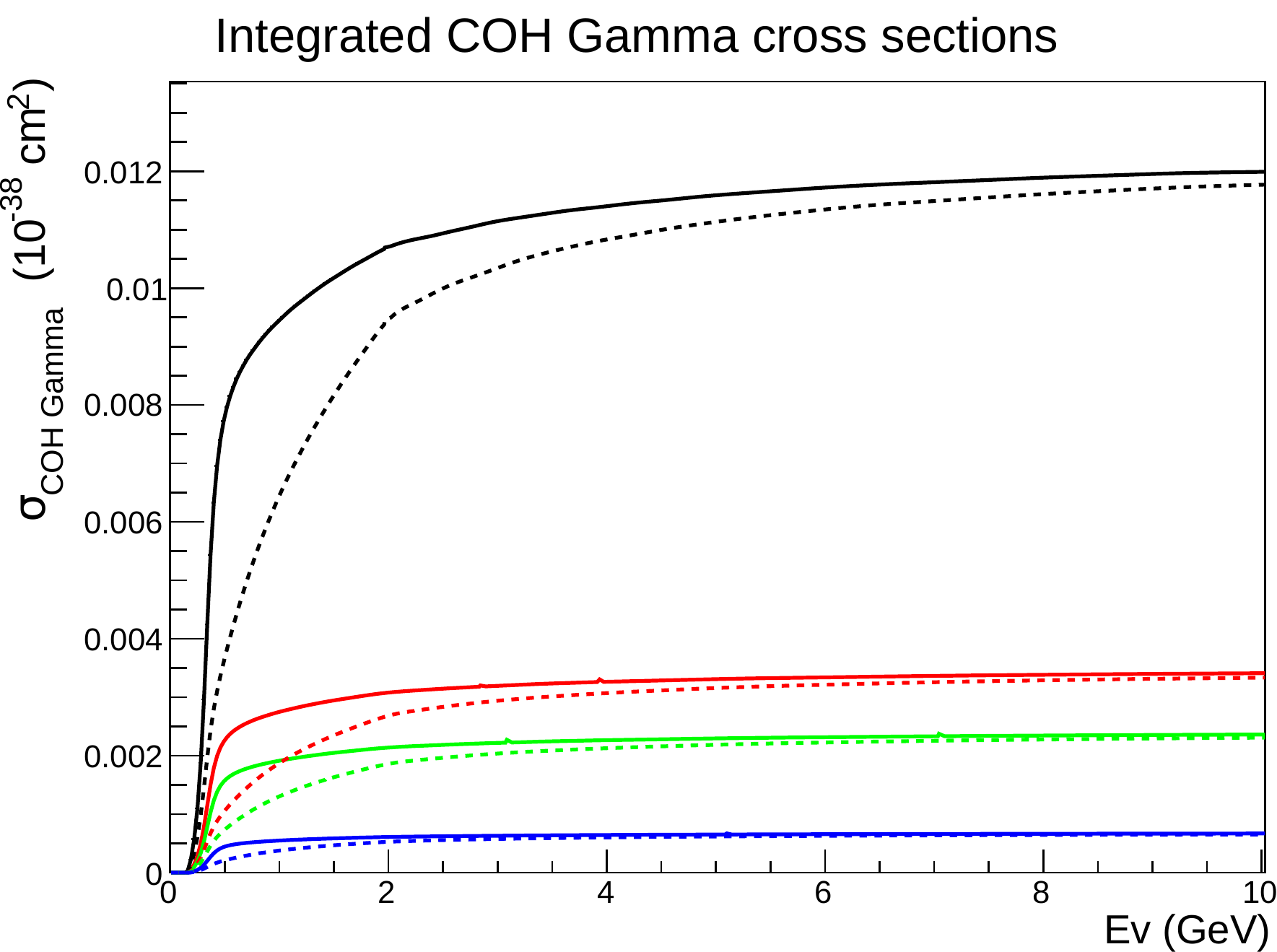}
    \caption{Integrated cross section of COH Gamma production from $\nu_\mu$ (continuous lines) and $\bar{\nu}_\mu$ (dotted lines) on $^{12}$C (blue), $^{40}$Ar (green), $^{56}$Fe (red) and $^{208}$Pb (black).}
    \label{fig:coh_gamma}
\end{figure}

In spite of its interest, the coherent excitation leading to a gamma production was missing and it has been included following the theoretical development of Ref.~\cite{coh_gamma_reference_paper} but introducing some simplifications to make event generation feasible \cite{EduardoSSThesis}.
Within a microscopic approach, the nuclear current is obtained by summing the contributions of all nucleons.
In this sum, the nucleon wave functions remain unchanged leading to nuclear density distributions and nuclear form factors. 
In the GENIE implementation, empirical parametrisations~\cite{DeVriesFormFactors} have been adopted for these form factors, adapted to any nucleus by means of interpolation. 
The total cross section for the coherent gamma production reaches a plateau for neutrino energies around 2~GeV, see \figurename~\ref{fig:coh_gamma} to see examples of integrated cross sections. 

The development is on target to be released in version v3.2.2 including only the dominant $\Delta$~resonance contribution. The contributions from heavier baryon resonances, which is a correction of the $10\%$ order sizable for (anti)neutrino energies above 1.5~GeV \cite{EduardoSSThesis} and potentially relevant for BSM searches at MINERvA,  will be released at a later stage.

\subsection{Final state interactions}
\label{sec:MEnu_fsi}


Final state interaction (FSI) models are a difficult but important part of the code.  Any hadrons produced in principal interaction models are then processed by one of the FSI codes.  As a result, any evidence of the principal interaction is masked as hadrons propagate through the residual nucleus.  
From the beginning, GENIE included the hA model which is a data-driven code that is fully reweightable.  
With version 3.0, the hN (full cascade very similar to NuWro and NEUT) was added.  With v3.2, we add the INCL++~\cite{Mancusi:2014eia,Cugnon:2016ghr} and Geant4 extended Bertini~\cite{Wright:2015xia} models.  These are added as libraries with appropriate interfaces.  Implementation mainly required a transfer of variables between codes so that the GENIE output would be as close as possible to the native FSI codes.  GENIE assumes the Monte Carlo method of choosing interactions as the particle propagates according to the mean free path which depends on  position and energy.  Both codes assume the hadron was in an incident beam and adjustments were required.  
For INCL, each hadron coming out of a principal interaction was separately propagated starting on-shell.  
Geant4 unfortunately has all particles interact and then normalizes to template hadron-nucleus cross sections.  This is incompatible with lepton production processes and was covered by having the hadrons use the same stepping process as hA and hN.  
Although this means the total reaction cross section is different than what the Geant4 code would calculate, the overall results in GENIE are a reasonable description of hadron-nucleus data.  
A separate article~\cite{Dytman:2021ohr} compares hA, hN, and INCL with other event generators for total reaction cross section and transparency. 
The conclusions there are similar to what is seen here.

Each of the newer models add significant capabilities.  
All of the newer codes are based on free hadron-nucleon cross sections with corrections to account for the surrounding nucleons. 
hN has medium corrections for both pions~\cite{Salcedo-Oset} and protons~\cite{Pandharipande:1992zz} and steps particles in space.  
In the released tunes, the nuclear model is consistent with the one used in the primary interactions. 
The original hA model tracks particles in the same way as hN except that it doesn't have pion medium corrections. INCL~\cite{Mancusi:2014eia,Cugnon:2016ghr} and Geant4~\cite{Wright:2015xia} use a series of shells at different radius, each having a custom depth.   
Therefore, both naturally include medium and binding energy corrections in a basic way.  

Each of the codes includes charge exchange and inelastic scattering, absorption (pions) and knockout (nucleons and kaons), and pion production processes.  Tracking of nucleons and pions is common to all codes, but INCL has no additional capabilities. 
hA and hN add $K^+$ and although Geant4 adds a host of additional particles, only kaons are presently enabled. 
Although, both hA and hN models have simple mechanisms to produce the well-known rise in nucleon yield at energies less than $\sim$20 MeV, INCL and Geant4 add the capability to simulate low energy compound nuclear processes and coalescence which adds light ions and photons to the final state.

\begin{figure}[htp]
\centering
\includegraphics[width=0.50\textwidth]{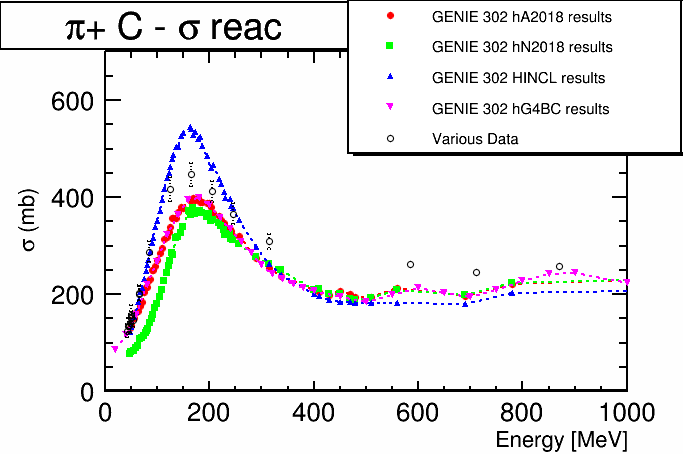}
\hfill
\includegraphics[width=0.46\textwidth]{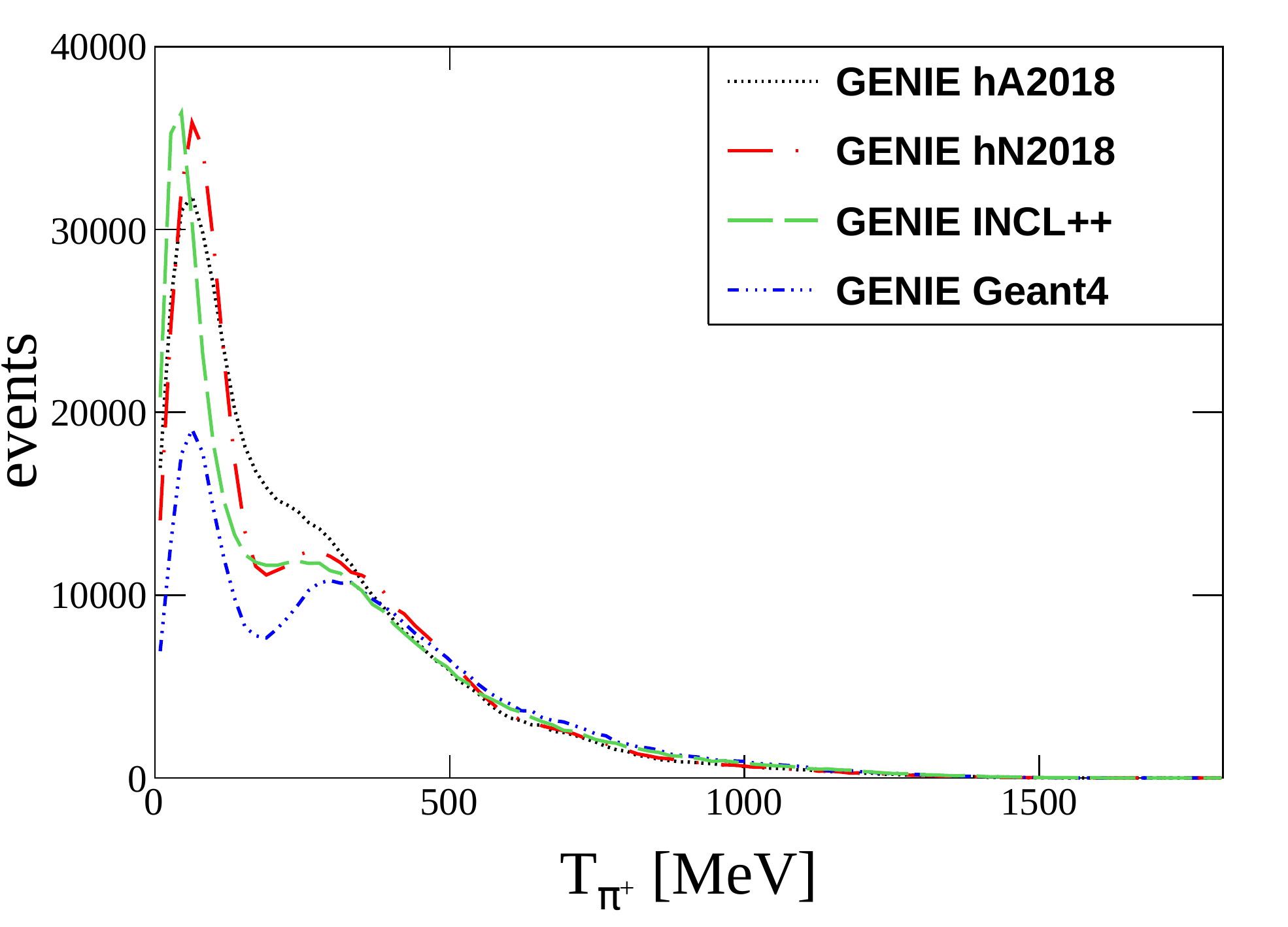}
\caption{Distributions for $\pi^+$, total $\pi^+$ reaction cross section for carbon (left) and inclusive $\pi^+$ kinetic energy distribution from 2 GeV $\nu_\mu\ {}^{40}$Ar (right).  In each case, results from all 4 models described in the text are shown.  }
\label{fig:fsi_pion}
\end{figure}

\begin{figure}[htp]
\centering
\includegraphics[width=0.51\textwidth]{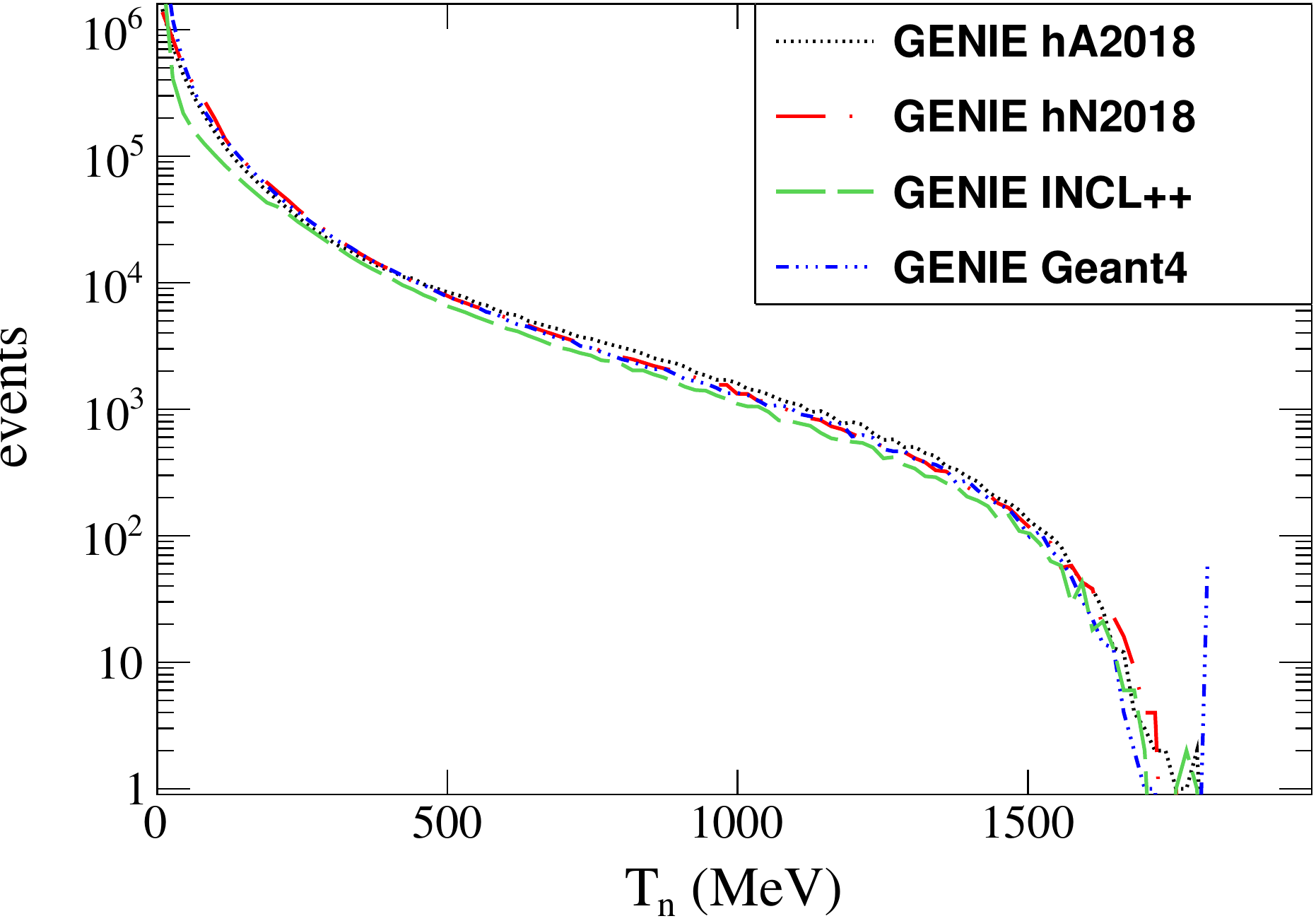}
\hfill
\includegraphics[width=0.48\textwidth]{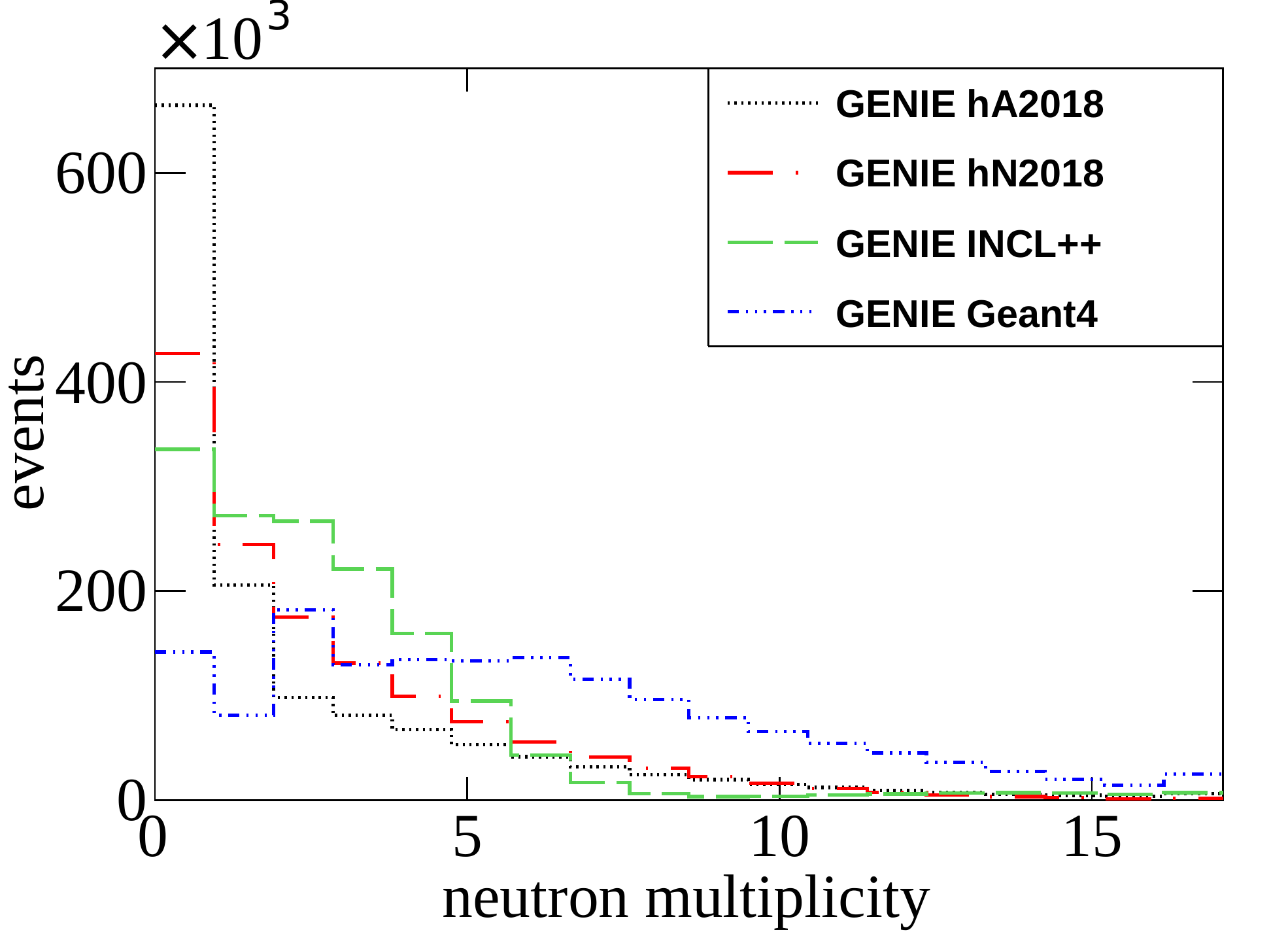}
\caption{Neutron distributions from a simulation of 2 GeV $\nu_\mu\ {}^{40}$Ar, kinetic energy (left) and multiplicity (right).  In each case, results from all 4 models described in the text are shown.}
\label{fig:fsi_neutron}
\end{figure}

\begin{figure}[htp]
\centering
\includegraphics[width=0.49\textwidth]{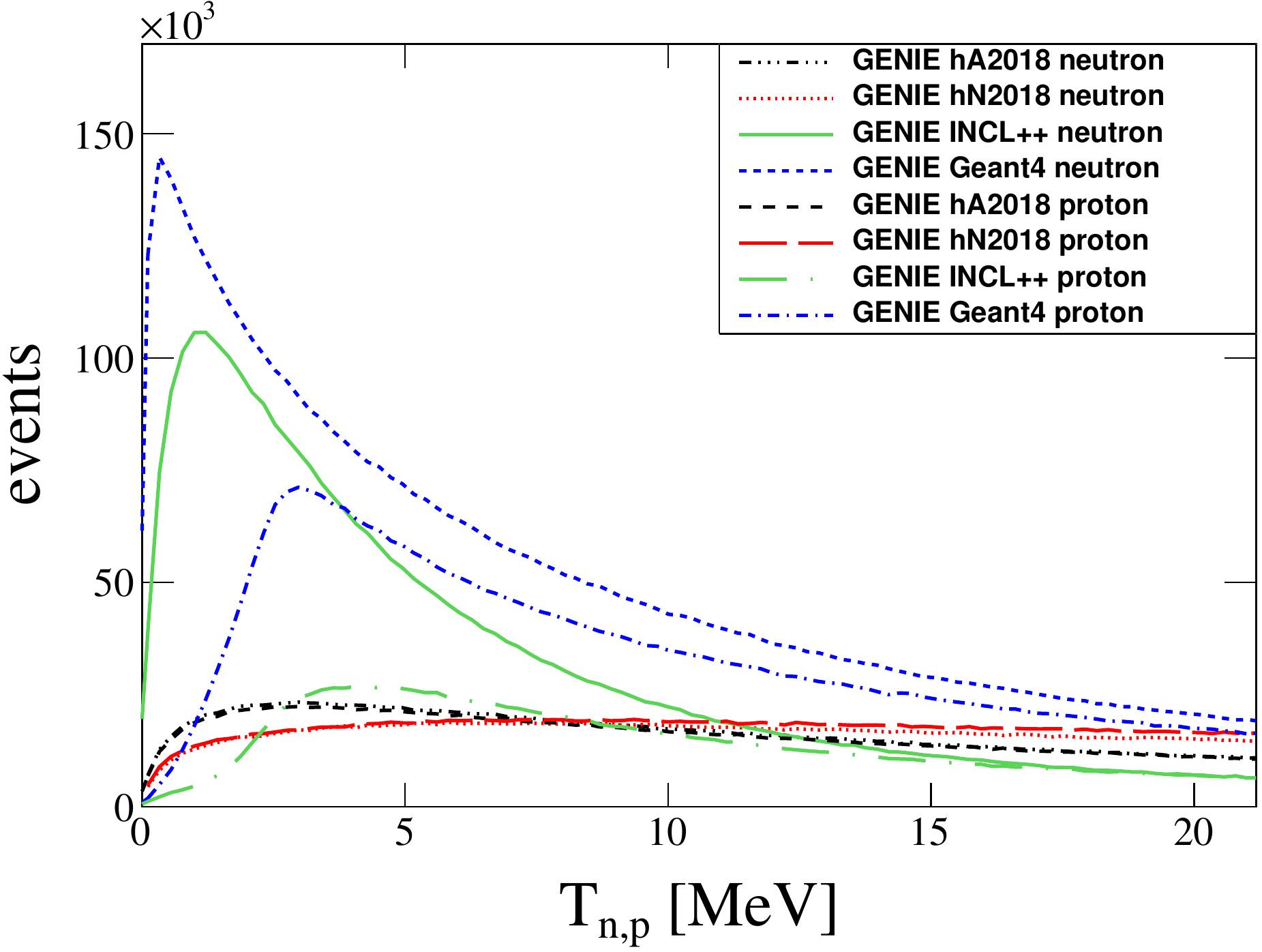}
\includegraphics[width=0.49\textwidth]{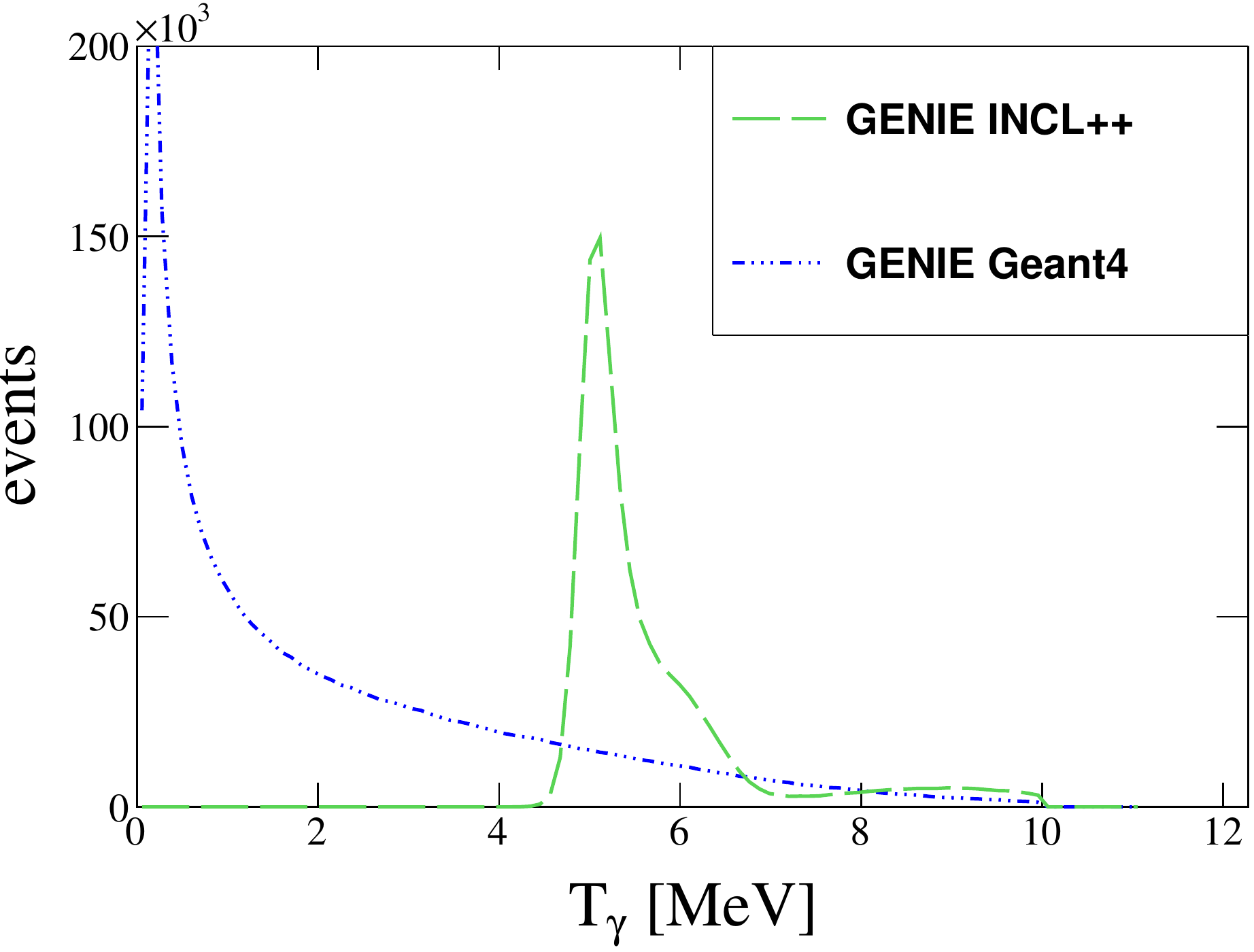}
\caption{Left: Neutron and proton kinetic energy distributions from a simulation of 2 GeV $\nu_\mu\ {}^{40}$Ar focusing on low energy responses, in each case, results from all 4 models described in the text are shown. 
Right: Photon kinetic energy distributions from a simulation of 2 GeV $\nu_\mu\ ^{40}$Ar focusing on low energy responses. 
Results from INCL++ and Geant4 models described in the text are shown. The gamma mode is not included in either hN or hA models. }
\label{fig:fsi_pncomp_and_gamma}
\end{figure}

Despite the wide range of approximations, all 4 models have similar general ability to describe data at higher energies (kinetic energy larger than $\sim$300 MeV), but show significant variations where nuclear effects are important. 
Fig.~\ref{fig:fsi_pion} left shows the total reaction cross section for $\pi^+$-carbon.  
More important properties of hadrons can be seen in a simulation of 2 GeV $\nu_\mu$-argon with all CC interactions enabled. 
Since each simulation included 2 million events using the same set of principal interactions, these distributions can be compared directly as cross sections. 
The pion kinetic energy spectrum (Fig.~\ref{fig:fsi_pion} right) is very similar for all models at energies above 300 MeV.  
However, there are significant differences around the peak of the $\Delta P_{33}$(1232) resonance, showing affects beyond what can be seen in Fig.~\ref{fig:fsi_pion}. 
Although, the kinetic energy distribution (Fig.~\ref{fig:fsi_neutron} left) is more similar among the models, strong deviations are seen at low energy.  
On the other hand, the neutron multiplicities (Fig.~\ref{fig:fsi_neutron} right) show wide variation according to model. 
Low energy proton and neutron kinetic energy spectra (Fig.~\ref{fig:fsi_pncomp_and_gamma} left) show a wide variation among the models because of the sensitivity to low energy modeling. This is one place where INCL and Geant4 show significant advantage.  Protons should be suppressed compared to neutrons at low energy by Coulomb effects and the newer models show this.  The other significant advantage is in the emission of photons and light ions.  The photon spectrum is shown in Fig.~\ref{fig:fsi_pncomp_and_gamma} (right). 
Both INCL and Geant4 use statistical models rather than conforming to specific states in the residual nuclei.

\subsection{HEDIS}
\label{sec:hedis}

Historically, the focus of neutrino interaction modelling in GENIE was the medium neutrino energy range (a hundred MeV up to a hundred GeV) relevant for atmospheric neutrino studies, as well as for studies of accelerator-made neutrinos both at short and long baseline experiments. GENIE has the mission to support the global experimental neutrino program and the emergence of the field of high-energy neutrino astronomy \cite{Aartsen:2013bka,Aartsen:2013jdh,Aartsen:2014gkd}, as well as the FASER$\nu$ \cite{Abreu:2020ddv} and SHiP \cite{Alekhin:2015byh} projects at CERN, generated the demand for accurate GENIE simulations of high energy neutrino interactions, beyond what was available through extrapolations of its model geared towards medium energies. 
To address this demand, a new {\em HEDIS} GENIE package
was created \cite{Garcia:2020jwr}, implementing high-energy cross section calculation and event generation modules. A new series of new CMC (\texttt{GHE19\_00a}, \texttt{GHE19\_00b}, \texttt{GHE19\_00c}, \texttt{GHE19\_00d}) using alternative HEDIS configurations were constructed. These new CMCs can be applied strictly for neutrino energies above 100 GeV, and have been validated up to $10^{9}$ GeV! Joining up the medium and high energy simulations into CMCs that span the full energy range will be the objective of a future development project.

\begin{figure}
    \centering
    \includegraphics[width=0.6\textwidth]{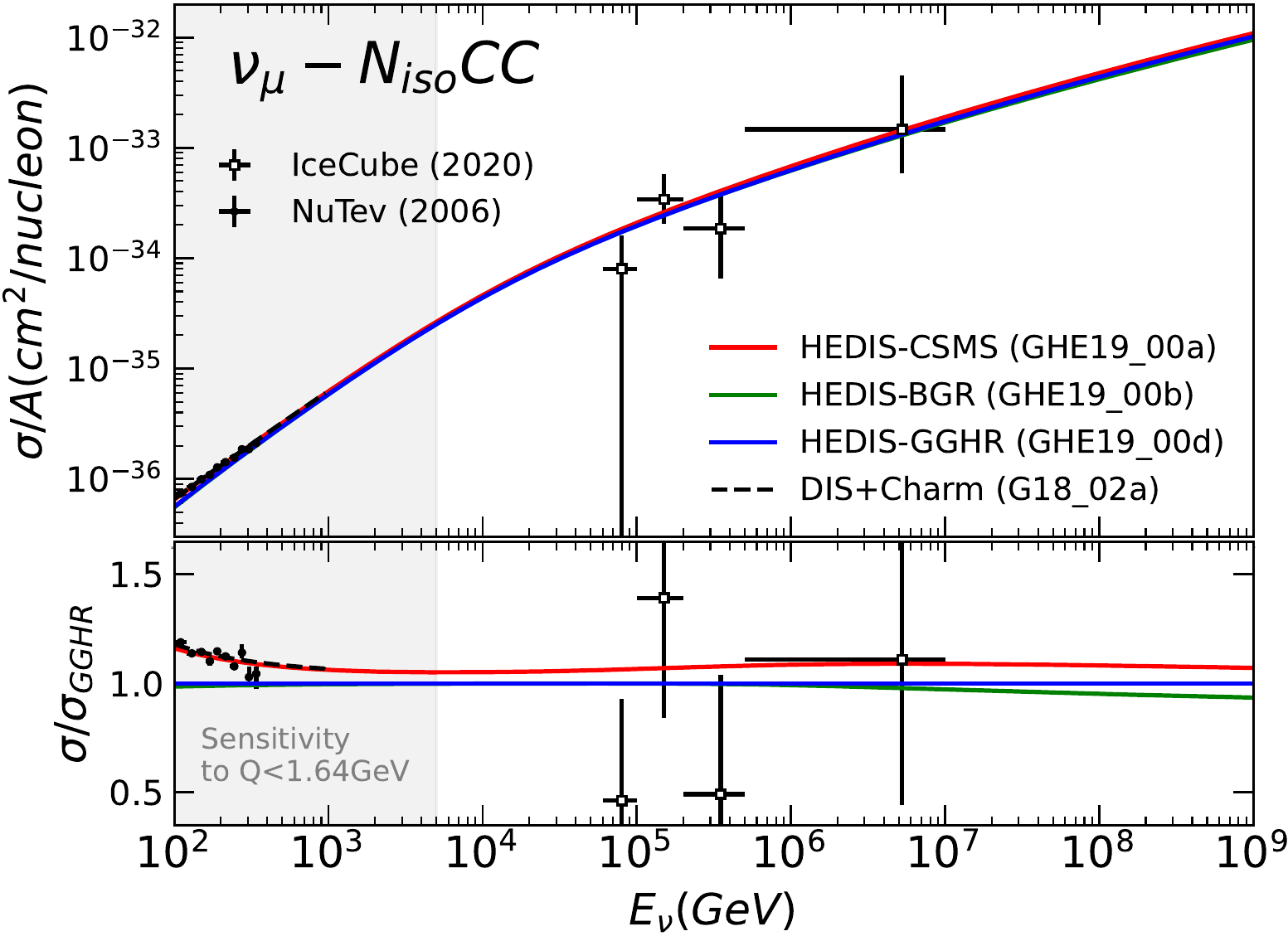}    
    \caption{Predictions of total cross section per nucleon for $\nu_\mu$ CC scattering assuming isoscalar target for three different models implemented in HEDIS: BGR18 \cite{Bertone:2018dse}, CMS11 \cite{CooperSarkar:2011pa} and GGHR20 \cite{Garcia:2020jwr}. Predictions from \texttt{G18\_02a} (assuming DIS and charm production) and NuTeV \cite{NuTeVxsec} and IceCube \cite{IceCubexsec} measurements are shown up to $E_{\nu}=1$~TeV for comparison. Grey area shows the region where low-momentum contributions ($Q<1.64$~GeV) are relevant.}
    \label{fig:xsec_hedis}
\end{figure}

The current HEDIS package includes several relevant scattering mechanisms relevant for high energy neutrinos. Where possible, changes were implemented through a new generalised interface for structure function calculations. Deep inelastic scattering (DIS) off nucleons which is modelled at NLO level. The calculation can incorporate sub-leading resonant DIS effects due to neutrino interactions with the photon field of the nucleon \cite{Garcia:2020jwr}. Generally, for DIS scattering off gluons and quarks, in the perturbative regime, the structure functions $F_{i}^{\nu N}$ factorise
in terms of process-dependent coefficients $C^{\nu}_{i,a}$ and
process-independent PDFs $f_a^N$ as follows

\begin{equation*}
\centering
F_{i}^{\nu N}(x,Q^2) = \sum_{a=g,q} \int_{x}^{1} 
 \frac{dz}{z} C^{\nu}_{i,a}\left(\frac{x}{z},Q^2\right) f_a^N\left(z,Q^2\right)   
\end{equation*}

\noindent where the coefficients $C^{\nu}_{i,a}$ can be computed in perturbation theory as a power expansion in the strong coupling constant $\alpha_s$. The evolution of PDFs is determined by the DGLAP equations. HEDIS has the option to account for the impact of logarithmic enhancements arising from high-energy gluon emissions (small-$x$ resummation) of coefficient functions and PDF evolution.
Structure functions are computed using the {\em APFEL} program \cite{Bertone:2013vaa}, whereas small-$x$ resummation through interface to {\em HELL} \cite{Bonvini:2018iwt}.
The main NLO DIS calculation implemented in HEDIS is the {\bf BGR18} \cite{Bertone:2018dse} model (\texttt{GHE19\_00a}).
In the BGR18 calculation, all inputs are at NLO accuracy. PDF sets are taken from the NNPDF3.1sx \cite{Ball:2017otu} global analysis of collider data, incorporating (through PDF reweighting) the impact of LHCb $D$-meson production in $pp$ collisions (small-$x$ PDF constraints beyond the kinematic range of HERA data) \cite{Aaij:2013mga,Aaij:2015bpa,Aaij:2016jht}. The calculation is using the FONLL scheme \cite{Forte:2010ta} to account for quark mass effects.
Nuclear corrections computed using the EPPS16 nPDFs \cite{Eskola:2016oht,Dulat:2015mca}.
In addition to BGR18, the {\bf CMS11} \cite{CooperSarkar:2011pa} and {\bf GGHR20} \cite{Garcia:2020jwr} NLO DIS calculations is also implemented in HEDIS for reference (\texttt{GHE19\_00b} and \texttt{GHE19\_00d}, respectively). 
Figure~\ref{fig:xsec_hedis} shows the prediction of the total cross section per nucleon for $\nu_\mu$ CC scattering for the three models described above. The relative rise of the CMS11 calculation in the low-$E$ region is due to the inclusion of low-momentum contributions ($1.0<Q<1.64$~GeV) which are absent in BGR18 and GGHR20. 
Besides deep inelastic scattering (DIS), HEDIS incorporates contributions from coherent scattering from the nucleus, which represents a substantial (5 -- 10\%) contribution for heavy nuclei) \cite{Garcia:2020jwr}.
Glashow scattering, which is simulated externally to HEDIS, is also incorporated in the high energy GENIE CMCs.
As part of the HEDIS development, an alternative interface to PYTHIA6 which originates from LEPTO \cite{Ingelman:1996mq} was installed in GENIE. A comparative analysis and consolidation of the two PYTHIA/GENIE interfaces will be the subject of a future development project.


\subsection{Coherent elastic neutrino-nucleus scattering}
\label{sec:cevns}

At energies below \SI{100}{\MeV}, neutrino interactions with complex nuclei are dominated by coherent elastic neutrino-nucleus scattering (CE$\nu$NS), a neutral-current reaction in which the final nucleus is left in its ground state. Since the only experimental signature is the small nuclear recoil kinetic energy, direct detection of CE$\nu$NS events is challenging.
Despite being anticipated theoretically several decades ago \cite{Freedman1974}, only two measurements \cite{Akimov2017,CEvNSAr} have been reported to date, both by the \mbox{COHERENT} experiment at Oak Ridge National Laboratory.
Due to the usefulness of precision CE$\nu$NS data for studying nuclear structure \cite{Ciuffoli2018,CEvNSStructure}, searching for physics beyond the Standard Model \cite{Billard2018}, and, perhaps, for monitoring reactors \cite{Bowen2020}, this process is the subject of increasing theoretical and experimental attention worldwide.

\begin{figure}
    \centering
    \includegraphics[width=0.50\textwidth]{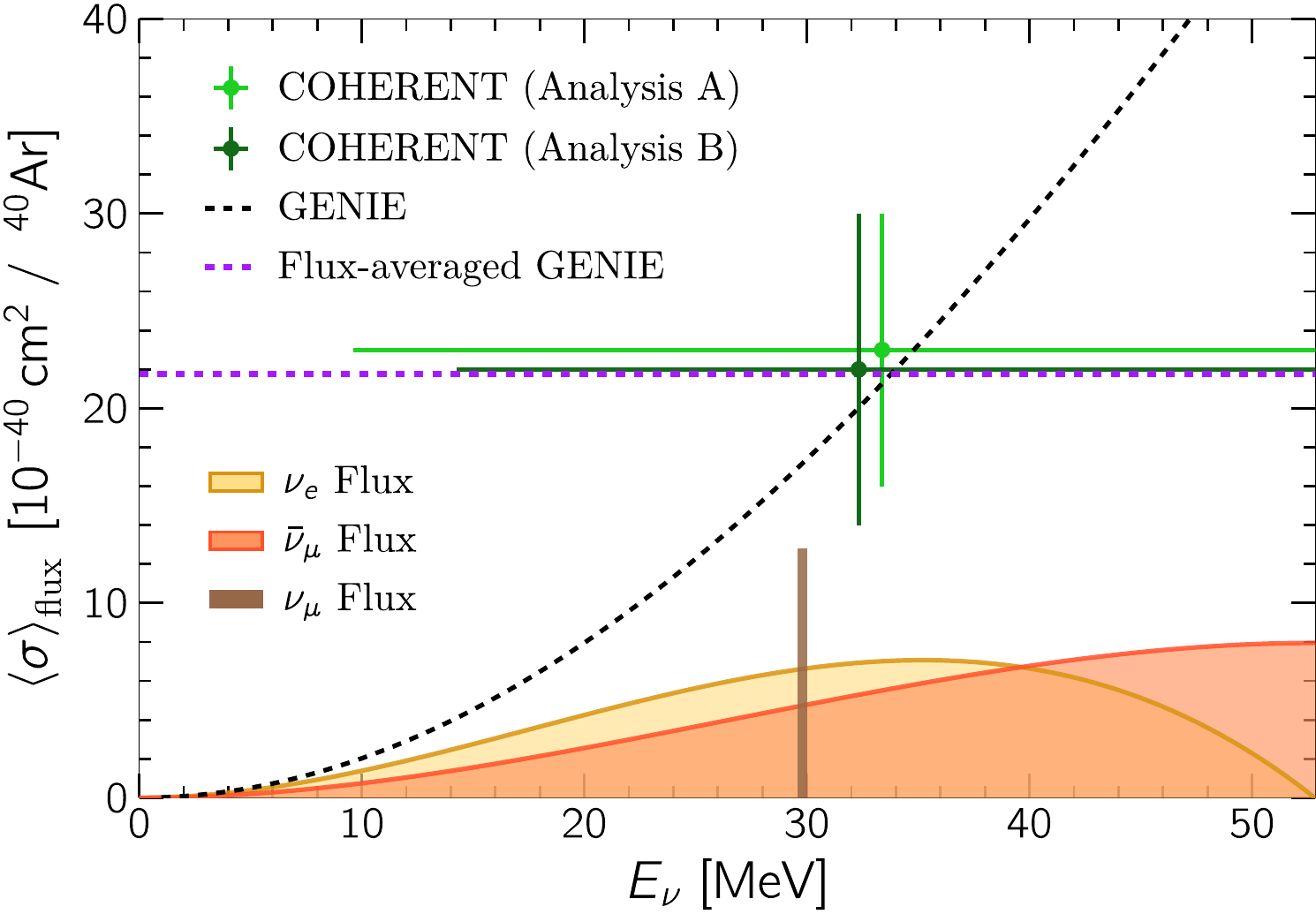}
    \caption{GENIE energy-dependent (dashed black) and flux-averaged (dashed violet) total cross section for coherent elastic neutrino-nucleus scattering on \isotope[40]{Ar} compared to data from the COHERENT experiment (green points). The neutrino flux spectra for the three participating neutrino species are also shown with arbitrary normalization.}
    \label{fig:cevns_data_mc}
\end{figure}

While proprietary codes are currently used by some experiments to simulate CE$\nu$NS, the GENIE v3 implementation represents the first realistic treatment of this process in a widely-distributed neutrino event generator.\footnote{The open-source MARLEY generator \cite{marleyCPC} provides a CE$\nu$NS model which relies on a rough approximation: the $Q^2$ dependence of the nuclear form factor is completely neglected.} 
In terms of the kinetic energy $T_A$ of the final nucleus, the CE$\nu$NS differential cross section is given by
\begin{equation*}
\frac{d\sigma}{dT_A} = \frac{G_F^2\,M}{4\pi} \, F^2\left(Q^2\right) \, \left[ 2 - \frac{2\,T_A}{E_\nu}
+ \frac{T_A^2 - M\,T_A}{E_\nu^2} \right],
\end{equation*}
where $M$ is the nuclear mass, $G_F$ is the Fermi constant, and $E_\nu$ is the incident neutrino energy in the laboratory frame. The nuclear form factor $F(Q^2)$ is sensitive to nuclear structure effects primarily through the neutron density distribution. 
The few-percent theoretical uncertainties on $F(Q^2)$ were recently studied for \isotope[40]{Ar} in Ref.~\cite{CEvNSStructure}. 
In GENIE, the form factor calculation currently used is that of Patton \textit{et al.} \cite{Patton2012}. 
The dashed black line in Fig.~\ref{fig:cevns_data_mc} shows the GENIE prediction for the CE$\nu$NS total cross section on \isotope[40]{Ar} as a function of neutrino energy. 
The flux-averaged prediction (dashed violet) is also shown for a pion and muon decay-at-rest neutrino source.
Excellent agreement is seen between the flux-averaged GENIE prediction and the recent COHERENT measurement \cite{CEvNSAr} (green points).

\subsection{New comprehensive model configurations and tunes}
\label{sec:MEnu_tunes}

The CMCs supported by the collaboration are natural evolutions of the GENIE v2 default configuration when a variety of roughly equal models for the same process are available. 
Care is taken to group compatible models together to maintain theoretical consistency and our goal of reproducing theoretical models accurately. 
CMCs are often grouped so that a full set of interactions is available for neutrino energies 100 MeV -- 1 TeV.

The first group of CMCs is historically motivated: it is based on the \emph{default} configuration in previous releases and simply provides updates for processes that were introduced later, like $\Lambda$ production or diffractive scattering from free protons. 
All these CMC IDs start with \texttt{G18\_01}.

The second family is an improvement of the first group in terms of the resonance model. 
 Specifically,  improvements of the Rein-Sehgal resonance models for CC and NC neutrino-production, as well as for CC and NC coherent production of mesons, were replaced with updated models by Berger and Sehgal~\cite{Berger:2007rq}.
 Their corresponding CMC are the \texttt{G18\_02}.

The third family, \texttt{G18\_10}, was constructed aiming to deliver the most up to date theoretical nuclear model simulations.
With respect to \texttt{G18\_02} CMCs, the $0\pi$ production models (Llewellyn Smith CC quasielastic  and GENIE's empirical multinucleon model) are replaced with implementations of the corresponding Valencia models by Nieves \textit{et al.}~\cite{Nieves}.
Within this group of models, the nuclear environment is modelled using a Local Fermi Gas, matching the inputs used for the published Valencia calculations. 
In the same groups belongs the CMCs that have the $0\pi$ production models based on the SuSAv2 approach, and they are labeled \texttt{G21\_11} (for neutrinos) and \texttt{GEM21\_11} (for electrons). 

Out of these main ideas, a number of CMCs can be constructed simply changing more detailed aspects like FSI or form factors. 
To indicate the FSI, one more letter is added to the CMC name: \texttt{a} for hA, \texttt{b} for hN, \texttt{c} for INCL, \texttt{d} for GEANT.

The complete tune names contain 2 additional fields that identify the tune performed using a CMC.
All fields must be specified, but the simplest choice is to use the same tune as in v2 by adding zeroes, e.g. \texttt{G18\_02a\_00\_000}.
More recent examples use the postfix \texttt{\_02\_11a} or \texttt{\_02\_11b} that denote a tune  against neutrino pion production data on protons and deuterium targets: specifically \texttt{\_02\_11b} identifies the tune described in~\cite{free_nucleon_tune}. 
Other notable examples are the hadronisation tunes described in~\cite{hadronisation_tune_paper} that have postfix \texttt{\_03\_330} or \texttt{\_03\_320} depending on the data used in the fit. 

\begin{figure}
    \centering
    \includegraphics[width=0.49\textwidth]{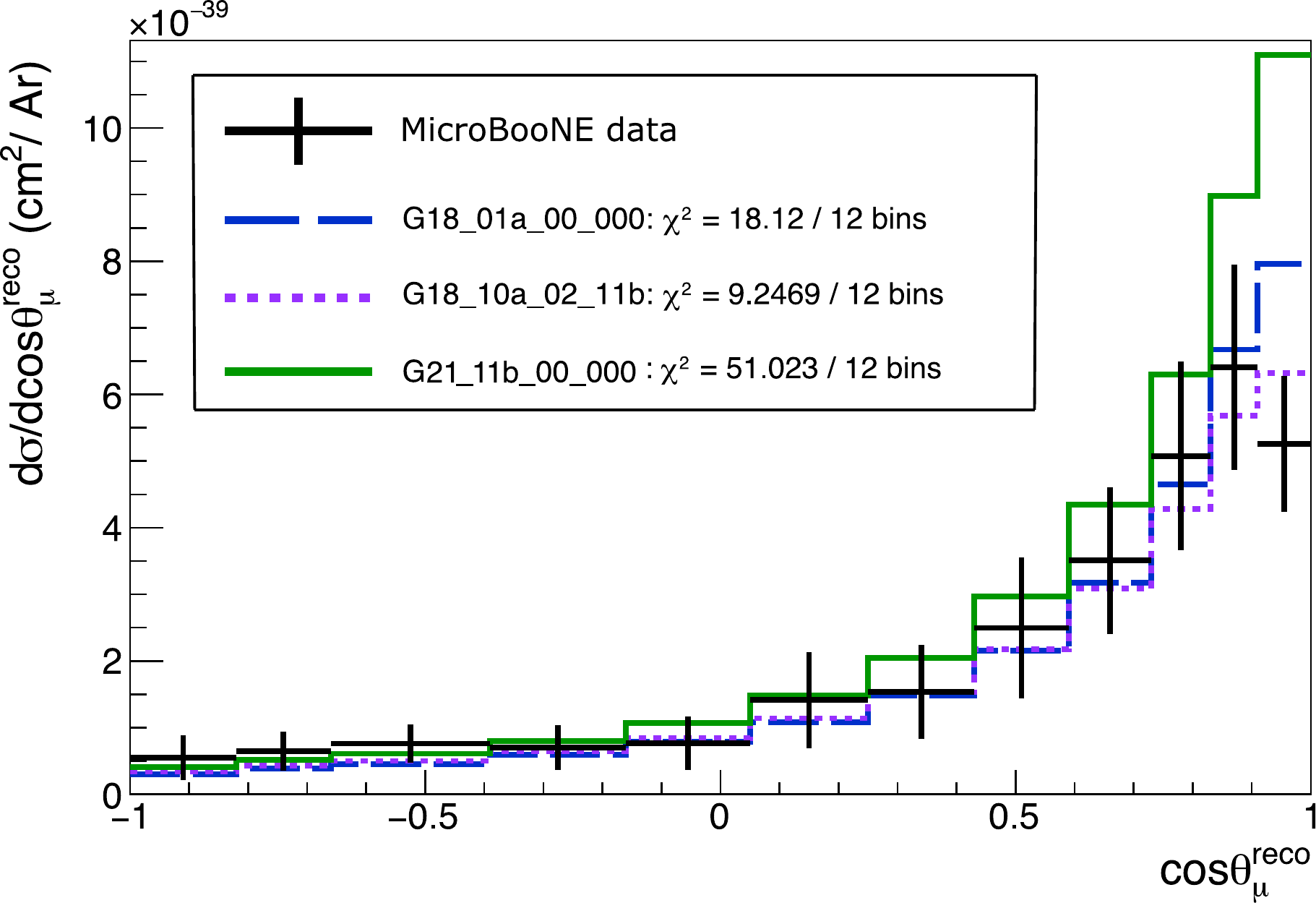}
    \hfill
    \includegraphics[width=0.49\textwidth]{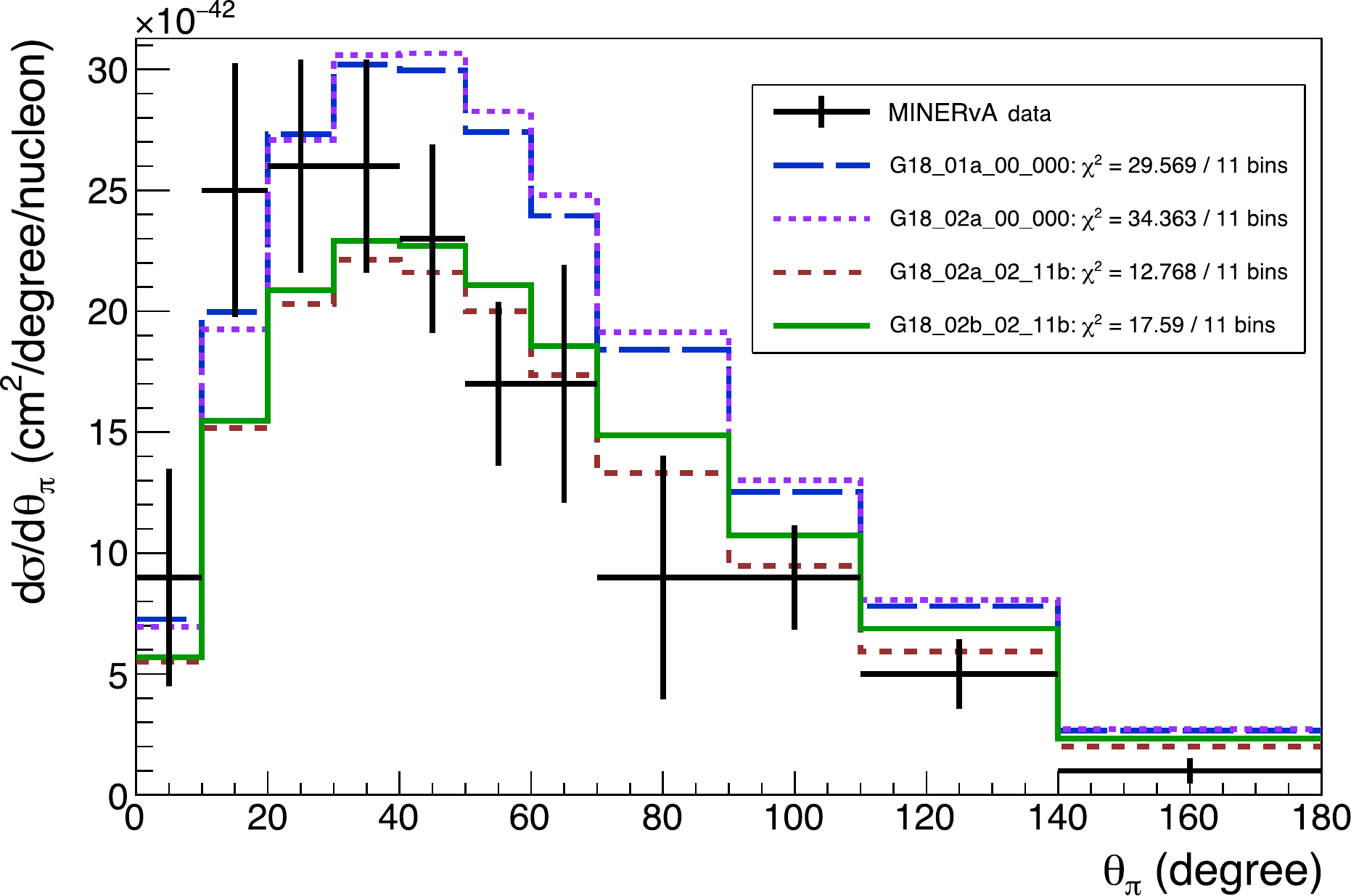}
    \caption{Comparisons of representative GENIE v3 CMC predictions to recent neutrino cross section measurements obtained by the MicroBooNE (left) and MINER$\nu$A (right) experiments.
    See the text for the dataset descriptions.
    }
    \label{fig:dataMCTunes}
\end{figure}

Figure~\ref{fig:dataMCTunes} compares recent neutrino cross section data to theoretical predictions generated using several different GENIE CMCs. 
The left panel shows the flux-averaged differential cross section obtained by MicroBooNE for the reconstructed muon scattering cosine in pionless $\nu_\mu$ CC events containing at least one final state proton \cite{uBooNECCNp}. The three CMCs compared to the data are
\texttt{G18\_01a\_00\_000} (dashed blue), \texttt{G18\_10a\_02\_11b} (dotted violet) and \texttt{G21\_11b\_00\_000} (solid green).
According to our naming scheme, the first is the historically motivated CMC that uses hA FSI, the second is the theory motivated tuned version, also using hA FSI, and the last is the latest implemented CMC that uses SuSAv2 with the hN FSI.
Substantially improved agreement is achieved by \texttt{G18\_10a\_02\_11b}  at forward angles, which is driven especially by the Valencia model's RPA-based treatment of long-range nucleon correlations. 
The right panel of Fig.~\ref{fig:dataMCTunes} shows a similar comparison to a measurement by the MINER$\nu$A Collaboration of single $\pi^{-}$ production in CC $\overline{\nu}_\mu$ scattering on hydrocarbon \cite{MINERvACCpiMinus}.
In this case, the choice of CMCs in the comparison emphasizes differences in the RES model and FSIs. 
\texttt{G18\_01a\_00\_000} (dashed blue) and \texttt{G18\_02a\_00\_000} (dotted violet) share the same hA model for FSIs but use the Rein-Sehgal~\cite{Rein:1980wg} and Berger-Sehgal~\cite{Berger:2007rq} treatments, respectively, to describe RES interactions.
Two additional CMCs are shown in which the Berger-Sehgal model is also used.
\texttt{G18\_02a\_02\_11b} employs exactly the same physics models as \texttt{G18\_02a\_00\_000}, but a number of parameters have been tuned based on fits to neutrino-nucleon scattering data \cite{free_nucleon_tune}. 
In \texttt{G18\_02b\_02\_11b}, the same tuned parameters are adopted, but the FSI model has been switched from hA to hN (see Sec.~\ref{sec:MEnu_fsi} for details). 
Although the other CMC differences in the comparison play some role, the improved agreement seen when using the tuned CMCs is the most significant effect.

\section{Beyond the Standard Model event generators in GENIE}
\label{sec:bsm}

Searches for physics Beyond the Standard Model (BSM) form an important pillar of the science program of current and future neutrino experiments. New detectors with unprecedented imaging capabilities, both massive ones at deep underground locations and smaller ones in near proximity to very intense proton and neutrino beams, create opportunities for expanding the sensitivity of established BSM searches and perform novel ones: Searches for nucleon decay, $n-\overline{n}$ oscillations, deviations from the SM neutrino trident rates, millicharged particles, dark neutrinos, light/boosted dark matter are, to name a few, some of improved or new BSM searches that will be carried out.
Standard neutrino interactions are a background to BSM searches and, therefore, it is
important to simulate both BSM and neutrino interactions in a common physics framework using, for example, common nuclear and intranuclear hadron transport modelling. 
GENIE supports these searches with a full implementation of four BSM generators: 
Generators for nucleon decay and $n-\overline{n}$ oscillations have been available in GENIE for several years. Recent additions, in GENIE v3 series, include the addition of a full Boosted Dark Matter (BDM) generator, and well as a first version of Dark Neutrino generator geared for low energy experiments.



\subsection{Boosted Dark Matter}
\label{sec:bsm_nnbar}

A BDM generator was made available in GENIE with the release of v3.0.0. The generator covers a extensive class of physics models described by the interaction Lagrangian
\begin{equation*}
    \mathcal{L}_{\text{int}} = g_{Z^\prime} Z^\prime_{\mu} J^\mu_{Z^\prime,\psi}
\end{equation*}
where
\begin{equation*}
    J^\mu_{Z^\prime,\psi} = 
     \overline{\psi} \gamma^\mu\left(Q_L^\psi P_L + Q_R^\psi P_R\right) \psi
\end{equation*}
and $\psi = {\chi, u, d, s, c, e}$.
The model is specified by charges $Q_{LR}^\psi$, the gauge coupling $g_{Z^\prime}$ and the masses of the dark matter particle $\chi$ and of the mediator $Z^\prime$.

A substantial upgrade of the BDM generator was deployed in GENIE 3.2.0, aligning it with the model described in Ref.~\cite{Berger:2018urf}. The upgrade allows for a broader set of particle physics models which may incorporate both vector and axial couplings as well as different isospin structures. The upgraded generator also enables simulations involving new probes (anti-dark-matter), a new target (scattering off electrons), and an improved model of the elastic scattering process which includes a pseudoscalar form factor.

\subsection{Dark neutrino generator}
\label{sec:bsm_darknu}
Dark neutrinos interactions arise from an extension of the SM Lagrangian adding a fourth neutrino flavour that mixes with the SM neutrinos~\cite{dark_neutrino_main_paper}. 
This extension can explain the low energy electromagnetic (EM) 
excess detected by short baseline experiments. 
These new dark neutrinos are relatively heavy (O(100) MeV). 
This extension comes with a new light neutral boson (lighter than the dark neutrino) that couples with both EM and weak charge, although the coupling with the weak charge is considered negligible as shown from model developers' fits. 
The new Lagrangian predicts a dark equivalent for every existing NC SM interaction.
These new interactions are not interfering with the normal interactions as they have a different final state as they produce the dark neutrino in the final state.
At the moment only the dominant interaction is implemented: the COH Dark (Quasi) elastic interaction, which is the dark equivalent of CE$\nu$NS. 
The implemented cross section was given to us by Pedro Machado, one of the model's authors. 
Details of the implementation, including the differential cross section, can be found in a GENIE public note \cite{docDB206}.

\begin{figure}
    \centering
    \includegraphics[width=0.60\textwidth]{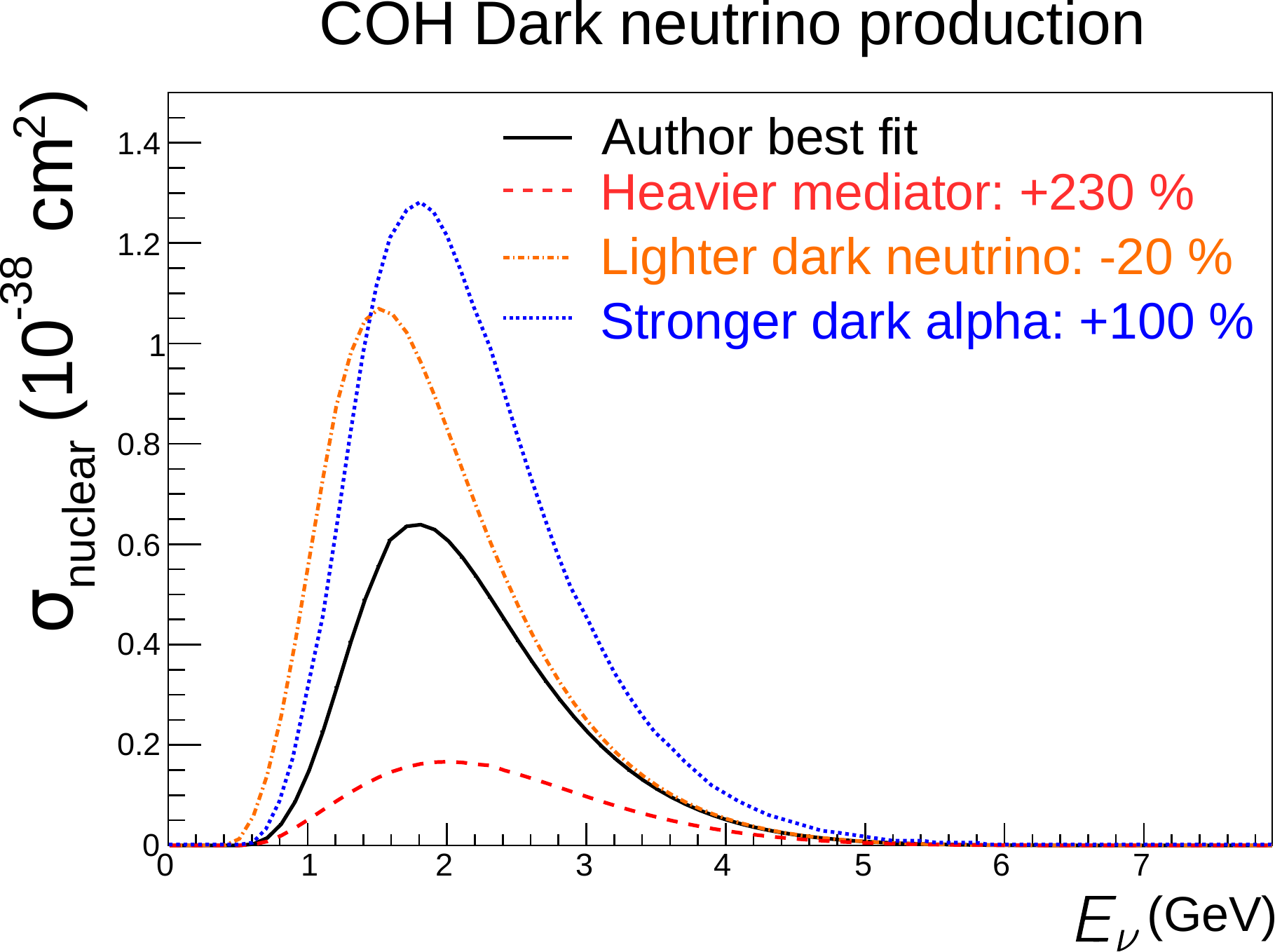}
    \caption{Integrated cross section for COH dark neutrino scattering from $\nu_\mu$ on Ar for different parameters. 
    The values used for the black curve are the one obtained from the MiniBooNE excess fit by the model authors in \cite{dark_neutrino_main_paper} and they are: $M_N=420$~MeV, $M_{Z_D}=30$~MeV and $\alpha_D=0.25$. Other curves varies one parameter at a time, according to the legend: the new values are selected purely for plotting purposes. }
    \label{fig:dark_neutrino_cross_sections}
\end{figure}

The model depends on several parameters: the masses of the dark particles ($M_N$ dark neutrino and $M_{Z_D}$ mediator), the neutrino mixing and the coupling between the dark boson and the EM charge. 
All these parameters affects the cross section: some just the intensity (mixings and $\alpha_D$) while the masses control the production threshold and the way that the cross section decreases as a function of $Q^2$.
Example of different parameters configurations can be seen in \figurename~\ref{fig:dark_neutrino_cross_sections}.

\section{Summary}
\label{sec:summary}

As neutrino physics continues to move into its precision era, the need for high-quality simulations of neutrino-nucleus scattering and related processes will only continue to grow.
GENIE has already been a standard tool in the field for many years, forming an indispensable part of many experiments' simulation workflows and offering a historical default model which has been widely tested against neutrino cross-section data. 
Major version 3 of GENIE builds upon this foundation with wide-ranging improvements to both physics modeling and to the technical machinery needed for daily use in experimental analyses.

A key technical addition in version 3 is the concept of a comprehensive model configuration, which allows multiple curated sets of compatible physics models to coexist in GENIE with a user-friendly means of switching between them. 
The new event library interface allows an external generator's physics models to be used within the GENIE framework while respecting the integrity of both codes. 
When these features are combined with a growing global analysis of neutrino scattering data and an advanced toolset for parameter tuning and uncertainty quantification, they form a state-of-the-art platform for meeting the simulation needs of current and future experiments.

The GENIE collaboration and external contributors continue to improve all aspects of the code's physics models, from the description of the nuclear target to hadronic final-state interactions, and including both standard and BSM processes. 
Improving the quality of GENIE's electron scattering mode and its consistency with the neutrino cross-section implementations has been an area of recent emphasis. 
The v3.2 release features various enhancements to GENIE's simulation capabilities for accelerator neutrinos, including entirely new model implementations for QE and $2p2h$ (SuSAv2), single pion production (MK), and intranuclear hadron transport (INCL and Geant4).
These core physics topics are complemented by substantial developments at lower (CEvNS) and higher (HEDIS) energies, reflecting the collaboration's mission to provide truly universal neutrino interaction modeling.


\section{Acknowledgements}
\label{sec:ackn}

We would like to thank the CC-IN2P3 Computing Center, as well as the Particle Physics Department at Rutherford Appleton Laboratory for providing computing resources and for their support.
This work, as well as the ongoing development of several other GENIE physics tunes was enabled through a PhD studentship funded by STFC through LIV.DAT,
the Liverpool Big Data Science Centre for Doctoral Training (project reference: 2021488).
The initial conceptual and prototyping work for the development of the GENIE / Professor interfaces, as well as for the development of the GENIE global analysis framework that, currently, underpins several analyses, was supported in part through an Associateship Award
by the Institute of Particle Physics Phenomenology, University of Durham.

This document was prepared by the GENIE collaboration using the resources of the Fermi National Accelerator Laboratory (Fermilab), a U.S.\ Department of Energy, Office of Science, HEP User Facility. Fermilab is managed by Fermi Research Alliance, LLC (FRA), acting under Contract No. DE-AC02-07CH11359.

The authors acknowledge support from US Department of Energy under contract No.~DE-SC0007914. 

A.P. acknowledges support from the Visiting Scholars Award
Program of the Universities Research Association. A.G. acknowledges support from the European Union’s H2020-MSCA Grant Agreement No. 101025085.
LAR has been supported by the Spanish Ministerio de Ciencia e Innovaci\'on and
the European Regional Development Fund (ERDF) under contract FIS2017-84038-C2-1-P, the EU STRONG-2020 project under the program H2020-INFRAIA-2018-1, grant agreement no. 824093 and by Generalitat Valenciana under contract PROMETEO/2020/023.

The GENIE code comes from the work of many people beyond the authors of this article and we are grateful to them.  
They include Josh Berger, Yanou Cui, Lina Necib, Yun-Tse Tsai, and Yue Zhao (boosted dark matter); Iker de Icaza and Pedro Machado (dark neutrino); Joe Johnston 
(Valencia QE model); Tomasz Golan
(Oset cascade model); Jarek Nowak (Berger Sehgal baryon resonance model);
Jackie Schwehr, Dan Cherdack, and Rik Gran
(Valencia $2p2h$ model); Jon Sensenig, Eduardo Saul-Sala, and Kathryn Sutton (coherent gamma); Chris Backhouse (event library); Stephen Dolan and Guillermo Megias (SuSAv2); and Dennis Wright and Makoto Asai (Geant4).


\printbibliography

\end{document}